\documentclass[12pt]{article}
\usepackage[utf8]{inputenc}
\usepackage{amsmath}
\usepackage{amsthm}
\usepackage{amsfonts}
\def\R {{\mathbb R}}

\usepackage{algorithm}
\usepackage[noend]{algpseudocode}
\usepackage{color}
\usepackage{lineno}
\usepackage{graphicx}
\usepackage{hyperref}
\usepackage{subcaption}
\usepackage[margin=1in]{geometry}
\usepackage[doublespacing]{setspace}
\usepackage{soul} 

\title{Bayesian Framework for Simultaneous Registration and Estimation of Noisy, Sparse and Fragmented Functional Data}
\author{James Matuk$^1$, Karthik Bharath$^2$, Oksana Chkrebtii$^1$, Sebastian Kurtek$^1$\\
$^1$Department of Statistics, The Ohio State University\\
$^2$School of Mathematical Sciences, University of Nottingham}
\date{}

\begin{document}

\maketitle

\linespread{1}
\begin{abstract}
In many applications, smooth processes generate data that is recorded under a variety of observation regimes, such as dense, sparse or fragmented observations that are often contaminated with error. The statistical goal of registering and estimating the individual underlying functions from discrete observations has thus far been mainly approached sequentially without formal uncertainty propagation, or in an application-specific manner. We propose a unified Bayesian framework for simultaneous registration and estimation, which is flexible enough to accommodate inference on individual functions under general observation regimes. Our ability to do this relies on the specification of strongly informative prior models over the amplitude component of function variability. We provide two strategies for this critical choice: a data-driven approach that defines an empirical basis for the amplitude subspace based on training data, and a shape-restricted approach when the relative location and number of local extrema is well-understood. The proposed methods build on elastic functional data analysis, which separately models amplitude and phase variability inherent in functional data. We emphasize the importance of uncertainty quantification and visualization of these two components as they provide complementary information about the estimated functions. We validate the framework using simulations and real applications to medical imaging and biometrics.

\end{abstract}

\newpage
\linespread{1.5}

\section{Introduction}

Functional Data Analysis (FDA) is an area of statistics in which the primary objects of interest are more naturally understood as functions rather than vectors \cite{ramsay_FDA,Ferraty:2006:NFD:1203444,srivastava_FASDA}. This perspective is advantageous in a wide range of application domains including biology, medicine, environmental science and engineering, where the underlying evolution of variables of interest is often smooth. Though the goals of FDA, including sample summarization and visualization, inference, and prediction, are similar to those of multivariate statistics, they are more challenging due to the inherent difficulty of working in infinite-dimensional function spaces. Importantly, functional data often exhibit two types of structural variability: amplitude variability, akin to differences in magnitude within variables in the multivariate setting, and phase or warping variability, related to differences in the timing of amplitude features, absent in the multivariate setting. Failure to account for these sources of variability in FDA can result in misleading summaries and biased inference \cite{marron2015}.

\subsection{Motivation}

The survey of FDA approaches in Section \ref{sec:relatedwork} highlights that existing methods frequently follow a sequence of estimation steps which perform the necessary tasks of (1) smoothing (projecting data to a lower-dimensional function space), (2) registration of smoothed functions (separation of amplitude and phase variability), and (3) summarization and/or inference. While conceptually straightforward, this pipeline generally suffers from two drawbacks: difficulty in formally propagating uncertainty between stages of estimation leading to over-confidence in the results, and lack of flexibility under different observation regimes. We argue that both of these issues can be resolved by a unified Bayesian inferential framework that performs smoothing, registration and inference simultaneously. Within this framework, data- or information-driven prior choices are necessary to extract meaningful observation-level information regardless of the collection protocol and/or issues with data quality, which often occur in practice.

\begin{figure}[!t]
\begin{center}
\begin{tabular}{cccc}
 \includegraphics[width = 1.46 in]{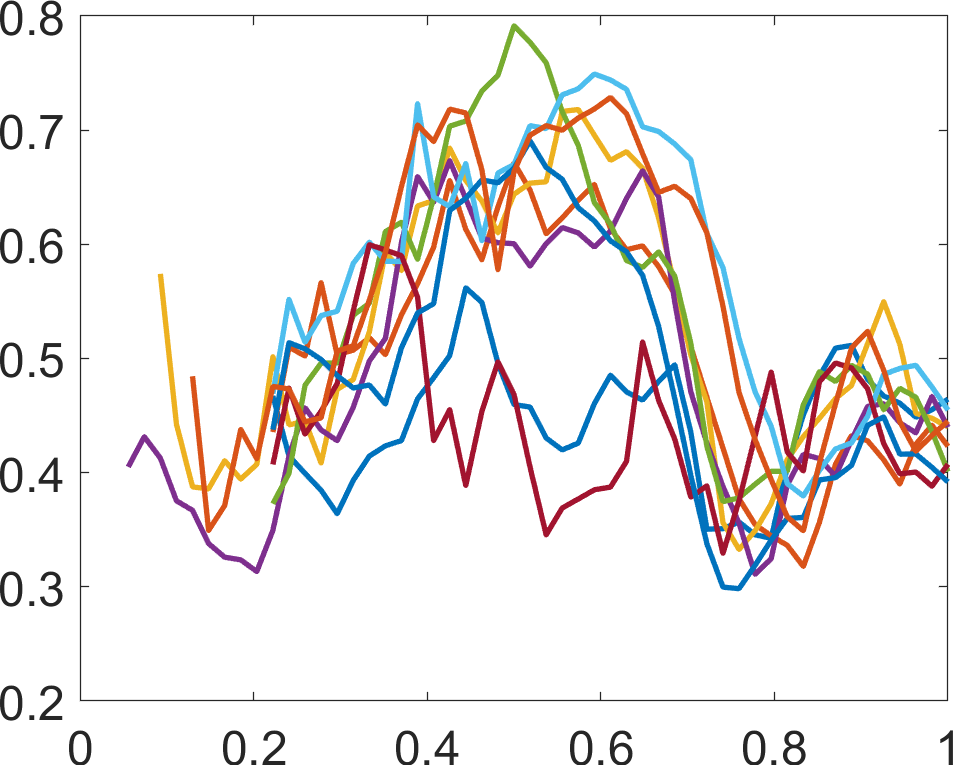} & \includegraphics[width = 1.46 in]{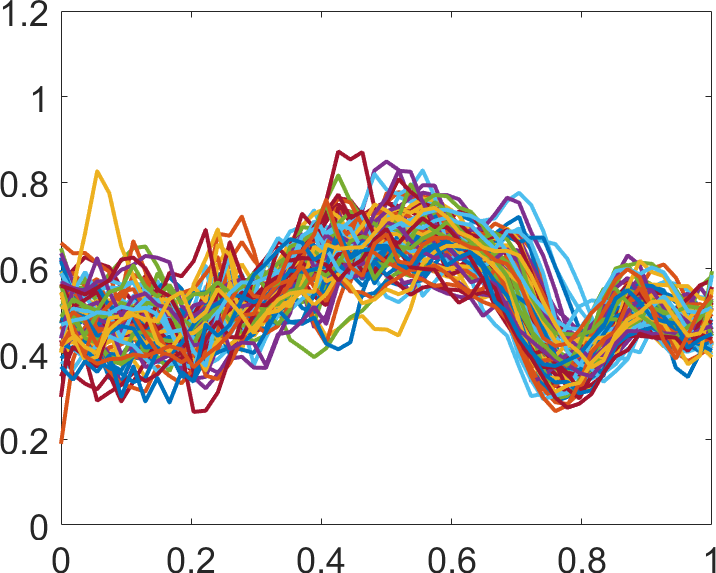} &  \includegraphics[width = 1.45 in]{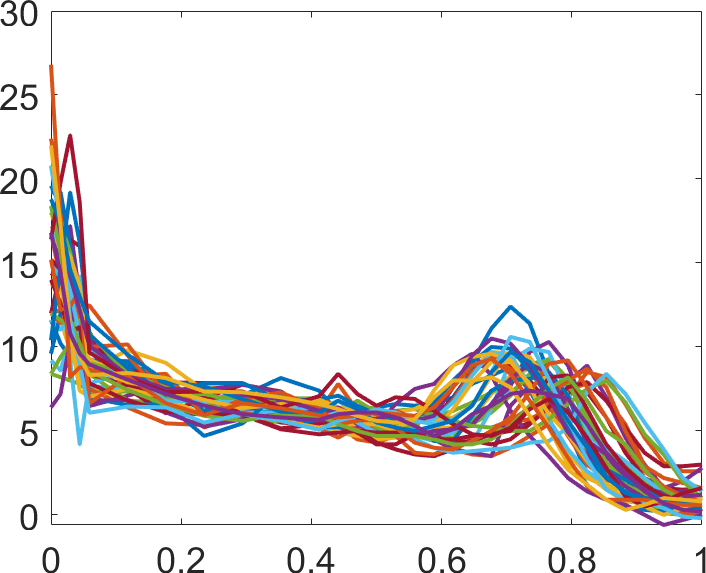} &\includegraphics[width = 1.46 in]{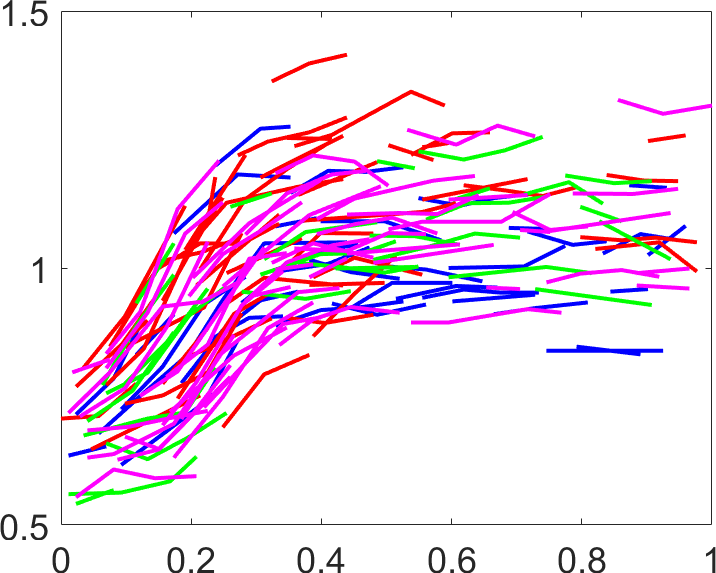}   \\
    (a) & (b) & (c)& (d) \\
    \end{tabular}
    \begin{tabular}{ccc}
     \includegraphics[width = 1.95 in]{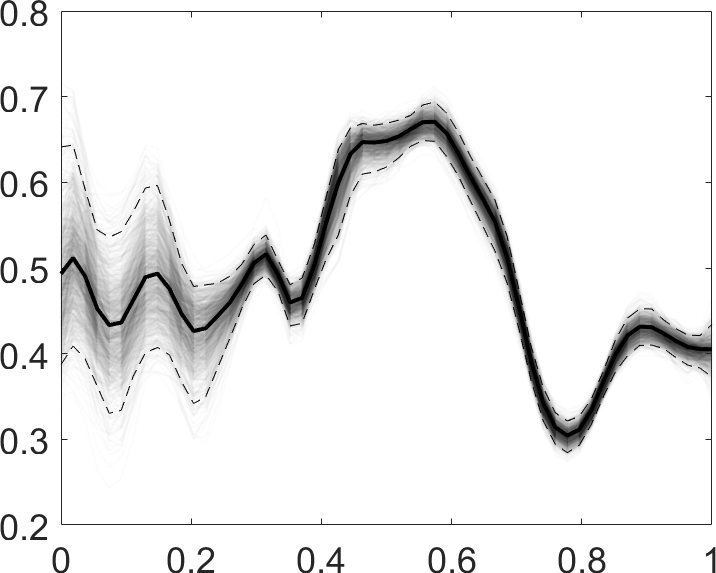} & \includegraphics[width = 1.58 in]{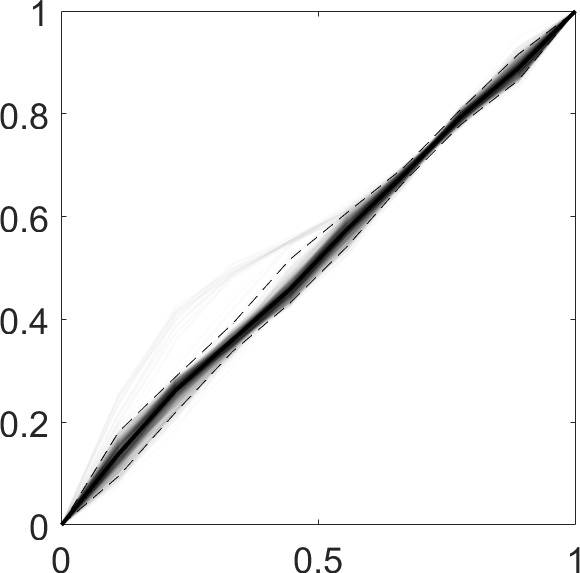}  & \includegraphics[width = 1.95 in]{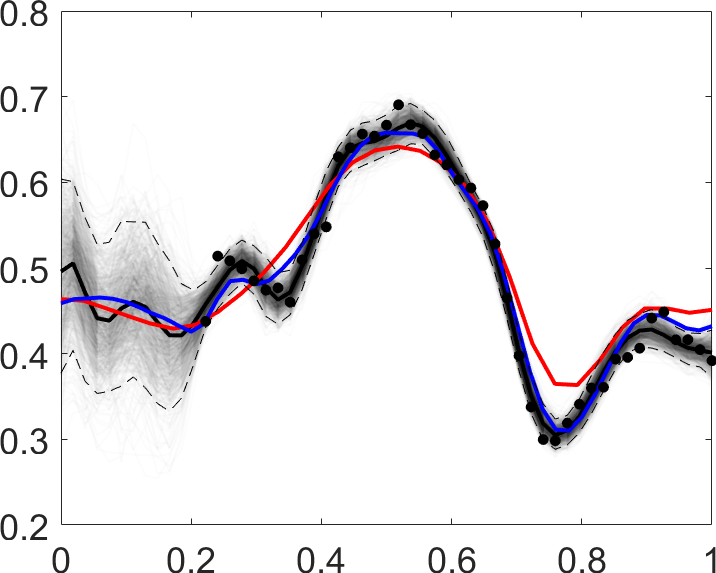} \\
     (e)&(f)&(g)\\
    \end{tabular}
    \caption{(a) Fragmented and (b) densely observed functional data from fractional anisotropy (FA) measurements. (c) Noisy growth velocity functions. (d) Fragmented and sparsely observed functions of bone mineral density (BMD). (e)-(g) Posterior summaries from the proposed Bayesian model for recovering a single FA function from (a) showing posterior uncertainty in the amplitude, phase, and composition of amplitude and phase, respectively. The blue and red functions in (g) correspond to estimates provided by existing methods.}
    \label{motivation_data}
    \end{center}
\end{figure}

In particular, functional data objects are observed on a finite subset of an interval $[a,b] \subset \mathbb{R}$, resulting in a variety of possible observational regimes:
(1) \textit{densely observed} functional data, wherein a large number of observations per function is available; (2) \textit{sparse} functional data, comprised of a small number of non-uniformly spaced observations per function; (3) \textit{fragmented} functional data, where observations of each function are unavailable over certain subsets of $[a,b]$. While scenario (1) is often assumed for modeling, scenarios (2) and (3) are common in biomedical and industrial applications. Additionally, in each of these scenarios, observations are frequently measured with noise or error. Figure \ref{motivation_data} illustrates such data and resulting inference on the underlying trajectory of a single observed function. Throughout this paper and in the figures, we transform the domain to the unit interval without loss of generality. Panels (a) and (b) show fractional anisotropy (FA) data, extracted from diffusion tensor magnetic resonance images (DT-MRIs), that are (a) fragmented early in the domain, or (b) densely observed. Panel (c) shows densely measured, but noisy, growth velocity curves for a group of children, and panel (d) shows both fragmented and sparse observations of bone mineral density (BMD). These examples vary in terms of the amount of sparsity, the degree of fragmentation, and the amount of noise. Panels (e)-(g) illustrate posterior uncertainty based on the proposed unified framework in different components (amplitude, warping, and their composition) of a smooth function underlying a single fragmented observation selected from panel (a). 
Posterior samples (grey lines) and pointwise $95\%$ credible intervals (dashed lines) illustrate the extent of posterior uncertainty, with two estimates generated by existing methods \cite{yao_pace,tang_registration} (PACE in red and WPACE in blue) provided for comparison.

Estimating the functions in such diverse observational regimes in the presence of amplitude and phase variation is a challenging problem. While the problem has been tackled for specific observation regimes, to the best of our knowledge, it has not been studied under a unified methodological framework. 
Such a framework should exploit the assumption that functional data share certain features, such as number of modes or inflection points. For example, in panels (a), (b) and (c) of Figure \ref{motivation_data}, the different functional observations contain similar numbers of modes, although their time of occurrence varies (phase variation) relative to a common state. The functions in panel (d) are generally increasing and have an inflection point whose location varies across observations. Moreover, such a framework should automatically provide a calibrated assessment of uncertainty and thus quality of registration under different observational regimes. We propose a Bayesian approach that satisfies these desiderata for simultaneous estimation and registration of functional data.


\subsection{Related work}\label{sec:relatedwork}

A popular approach to estimate underlying functions from discrete observations is penalized least squares (PLS \cite{wahba_spline}). Coefficients of a truncated basis expansion are estimated via least squares subject to an application-specific penalty, which enforces a soft constraint on some features of the function such as ``wiggliness'' as measured by total concavity \cite{ramsay_FDA,srivastava_FASDA}. 
Subsequently, the phase and amplitude variation of the estimated functions can be extracted via landmark, metric- or model- based registration procedures \cite{ramsay_FDA,srivastava_FASDA,kneip_registration,lu_registration,gerda_2010}. A drawback is the lack of a systematic mechanism to propagate uncertainty between the stages of such procedures. While \cite{lu_registration} propose simultaneous smoothing and registration within a Bayesian framework, the model pools information from all available sample elements to estimate a common underlying state, or template, rather than the underlying functions, which are often of interest, but cannot be estimated individually in data-poor observation settings.


Certain observation regimes pose further difficulties. Under a low signal-to-noise ratio, individual estimation of functions via PLS tends to overlook population-level features, resulting in misleading uncertainty estimates in subsequent analyses. To address this problem, methods have been developed to simultaneously infer the functions underlying the observations and population quantities such as mean and covariance functions. Mixed model approaches have been effective in modeling discrete observations with a fixed population mean function, random subject-specific functions and pointwise error terms \cite{shi_mixedmodel,yang_mixedmodel,rice_mixed}. However, none of these approaches consider phase variability as part of the model. In contrast, Raket et al. \cite{raket_mixed} define a mixed effects model that partially accounts for phase variability by assuming a random subject-specific warping of the population mean. Descary and Panaretos \cite{descary_matrix} take an entirely different approach and model both small- and large-scale variation instead of considering small-scale variation as noise. Underlying functions are modeled as a mixture of smooth and rough components and are estimated from discrete, noisy data. 

Shape-invariant models \cite{lawton_sim,fu_registration,telesca_registration} jointly estimate a population template function and subject-specific scaling factors, translations and warpings from discrete, noisy observations. However, such models are often too rigid for inference of individual functions that may exhibit more complex subject-specific variation which cannot be captured by translating, scaling and warping a single, fixed template function. A related set of approaches consider shape-constrained estimation \cite{wheeler_spline}, wherein additional structure is built into the model, such as the number of peaks and valleys, to constrain the resulting shape of the estimated functions. This idea was also recently used in the context of density estimation by Dasgupta et al. \cite{Dasgupta_shape} to constrain the number of modes. Methods to estimate functional data that is sparsely-observed with pointwise noise have also been developed. We adopt the informal convention that a functional observation is sparsely recorded if the number of observations for a subject is orders of magnitude smaller than the number of subjects in the dataset, as is common in longitudinal studies \cite{wang_fda}.
The proposal in \cite{rice_mixed} and \cite{telesca_registration} consists of fitting mixed models and shape-invariant models to sparse data, respectively. Principal Analysis through Conditional Expectation (PACE), a framework for inference on sparse functional data proposed in \cite{yao_pace}, pools discrete observations across all sample elements to estimate mean and covariance functions via smoothing and functional Principal Component Analysis (fPCA).
%

Classification of fragmented functional data was studied in \cite{delaigle_BMD,james_BMD}, without focusing on estimation or registration. Estimation of the individual functions, along with the population mean and covariance, for fragmented functional data was studied in \cite{DH} and \cite{DK} under fragmentary regimes distinguished by their assumptions on the coverage of observation intervals. While these works implicitly assumed a missing-at-random mechanism for the unobserved parts of the function, recent work in \cite{DL} carried out the estimation tasks under a missing-not-at-random assumption.

\subsection{Contributions and Organization}

The issues raised above may be addressed by simultaneously combining coherent uncertainty propagation between smoothing, registration and inference, while automatically informing both the phase and amplitude structure of individual functions. In particular, we argue that an inferential framework under a general observation regime must include an automatic method to appropriately restrict the model space \cite{raket_mixed,wheeler_spline}, whether based on a physical understanding of the underlying process or on previously observed data. Bayesian inference is a natural framework for introducing such hard and soft constraints through the choice of a prior distribution. We focus on two automatic approaches for prior specification on subject-specific amplitude components to carry out this restriction:
\begin{enumerate}
    \item When densely-observed data is available, a data-driven prior for inference on sparsely observed and/or fragmented samples drawn from the same population (e.g., Figure \ref{motivation_data}(a)-(b)) is designed to capture important modes of amplitude variation.
    \item  When concrete physical understanding about the data-generating mechanism is available, such as the distribution of peaks and valleys (e.g., Figure \ref{motivation_data}(c)-(d)), we leverage a prior that informs the shape and smoothness by controlling the number of extrema.
\end{enumerate}
We further model phase variation via a recently proposed point process-based prior on warping functions compatible with the discrete nature of sparse and fragmented data \cite{bharath_phase}.

While Bayesian models for functional data with phase variation have been considered under the sparse and dense observational regimes (e.g., \cite{telesca_registration,ZSGM}), we are unaware of any work that deals with estimation for fragmented functional data in the presence of phase variation. Moreover, the proposed model for handling amplitude and phase variation in the dense and sparse regimes adds flexibility relative to current approaches, by allowing for subject-specific templates rather than assuming a common shape template.

The remainder of this paper is organized as follows. Section \ref{phase_amplitude} describes an approach for prior modeling of amplitude and phase variability in densely observed functional data, which is used to inform the Bayesian model described in Section \ref{models}; this enables subject-specific inference under general observation settings. We validate the proposed framework on simulated and real data in Section \ref{examples}, and close with a brief discussion and some directions for future work in Section \ref{conclusion}. The Supplementary Materials include (1) detailed statistical analysis of the complete FA functions, (2) an additional simulation for the shape-restricted amplitude prior model, (3) an additional real data example that considers estimation of CD4 cell count functions for HIV patients, (4) the Markov chain Monte Carlo (MCMC) algorithm used to sample from the posterior distribution and other implementation details, and (5) MCMC diagnostics.


\section{Phase-Amplitude Separation via EFDA}
\label{phase_amplitude}

Functional samples from a population often share common features, such as the number and relative location of peaks and valleys. Differences in magnitude and location of these features along the domain are commonly referred to as amplitude and phase variability, respectively, as formalized in \cite{ramsay_FDA,srivastava_FASDA,marron2015}. It has been shown in multiple places \cite{srivastava_FASDA,srivastava_registration,tucker_pca} that modeling functional data without appropriately accounting for phase variability can result in inaccurate descriptive statistics and biased or misleading inference. Decoupling these two sources of variation is known as registration or alignment of functional data. For a sample of densely-observed functions, $f_i:[0,1]\to \R,\ i=1,\ldots,k$, a registration procedure yields corresponding amplitude functions $\tilde{f}_i:[0,1]\to \R$ and phase functions $\gamma_i$. 
Their composition $f_i = \tilde f_i \circ \gamma_i,\ i = 1,\ldots,k$ reconstructs the original functions uniquely.
A formal registration procedure must thus define a representation space of phase (these can vary in flexibility from linear to diffeomorphic transforms) and an appropriate optimality criterion.

To perform simultaneous smoothing and registration coupled with informative shape restriction, we choose a warping functional form and optimality criterion based on the elastic functional data analysis (EFDA) framework of Srivastava et al. \cite{srivastava_registration}. Indeed, its ability to define amplitude purely in terms of the shape of a function, that is, the number and (relative) heights of peaks and valleys independent of their locations on $[0,1]$, is consistent with our goal of semi-parametric shape restriction. Phase variability is modeled via smooth, monotone transformations of $[0,1]$: $\Gamma = \{\gamma:[0,1]\rightarrow[0,1]\, | \, \gamma(0) = 0, \gamma(1) = 1, \dot{\gamma}>0\}$, where $\dot{\gamma}$ is the time derivative. Though most other methods use the $\mathbb{L}^2$ metric as the optimality criterion for registration, this choice suffers from major deficiencies including the pinching effect (distorted amplitude) and asymmetry in registration (ill-defined amplitude) \cite{srivastava_FASDA,marron2015}. This is because the $\mathbb{L}^2$ metric is not invariant to simultaneous warping of functions, making its use inconsistent with the goal of registration. Instead, EFDA uses an extension of the Fisher-Rao Riemannian metric as a foundation for registration and statistical modeling \cite{srivastava_registration}; we omit its formal definition here for brevity. While this choice has useful mathematical properties, including the critical invariance to identical warping, it is difficult to use directly. However, the simple transformation described below reduces this metric to the standard $\mathbb{L}^2$ metric, facilitating simplified computation.

Let $\mathcal{F}=\{f:[0,1]\to\mathbb{R}\,|\,f\text{ is absolutely continuous}\}$ denote the function space of interest. For any $f\in\mathcal{F}$, define its square-root velocity function (SRVF) using the mapping
$$
Q:\mathcal{F}\to\mathbb{L}^2([0,1],\mathbb{R}), \quad Q(f)= \text{sign}(\dot{f})\sqrt{|\dot{f}|}=:q\thinspace.
$$
The space of SRVFs corresponding to $\mathcal{F}$ is denoted as $\mathcal{Q}$. Because the SRVF is invertible up to a translation, the original function can be reconstructed from its SRVF using $f(t) = Q^{-1}(q,f(0)) = f(0) + \int_0^t q(s)|q(s)|ds$. The extended Fisher-Rao (eFR) distance on $\mathcal{F}$ simplifies to the $\mathbb{L}^2$ distance on $\mathcal{Q}$: the eFR distance between two functions $f_1,\ f_2\in\mathcal{F}$ is computed as $d_{eFR}(f_1,f_2):= \|q_1 - q_2 \|=\left[\int_0^1(q_1(t)-q_2(t))^2dt\right]^{1/2}$.

EFDA defines the amplitude of a function $f$ as an equivalence class $[f]:=\{f \circ \gamma \, | \, \gamma \in \Gamma\}$ under the equivalence relation $f \sim g$ if there exists a warping $\gamma \in \Gamma$ such that $f \circ \gamma=g$. Equivalently, on the SRVF space $\mathcal{Q}$ we have $[q] := \{(q, \gamma) \,|\, \gamma \in \Gamma \}$, where $(q,\gamma) = (q\circ\gamma) \sqrt{\dot{\gamma}}$ is the corresponding action of $\Gamma$ on $\mathcal{Q}$ under the SRVF map $f \mapsto q$. The set of equivalence classes forms a partition of $\mathcal{Q}$, and is referred to as the quotient space $\mathcal{Q}/\Gamma$, i.e., $\mathcal{Q}/\Gamma$ defines the amplitude space. Therefore, registration under this framework requires the determination of an average or mean equivalence class, and alignment of all functions to one of its elements. Let $f_1,\ldots, f_k$ denote a sample of densely-observed functions, and $q_1,\ldots, q_k$ their corresponding SRVFs. The sample mean equivalence class is taken to be the Karcher mean: $[\hat{\mu}_q] = \underset{[q]\in \mathcal{Q}/\Gamma}{\text{argmin}} \sum_{i = 1}^k d([q],[q_i])^2=\underset{[q]\in\mathcal{Q}/\Gamma}{\text{argmin}} \sum_{i = 1}^k\underset{\gamma \in \Gamma}{\text{min}}\|q - (q_i,\gamma)\|^2$. To ensure identifiability, a representative element of the mean equivalence class, $\hat{\mu}_q \in [\hat{\mu}_q]$, is chosen such that the average of the optimal warpings of all functions to $\mu_q$ is the identity warping $\gamma_{id}(t)=t$. In addition to the mean amplitude function, $\hat{\mu}_f = Q^{-1}(\hat{\mu}_q,\bar f(0))$ where $\bar f(0)=\frac{1}{k}\sum_{i=1}^k f_i(0)$, this registration procedure also yields (1) phase functions, $\gamma^*_i=\underset{\gamma \in \Gamma}{\text{argmin}}\|\hat{\mu}_q - (q_i,\gamma)\|,\ i = 1,\ldots,k$, (2) registered SRVFs, $(q_i,\gamma^*_i),\ i = 1,\ldots,k$, and (3) amplitude functions, $\tilde f_i = Q^{-1}((q_i,\gamma^*_i),f_i(0)),\ i = 1,\ldots,k$.

Functional Principal Component Analysis (fPCA) is a decomposition of the total sample variation into orthogonal modes of variability. We propose to perform fPCA on amplitudes to construct informative empirical priors over this component of variation. Also called vertical fPCA, it was first used within the EFDA framework by \cite{tucker_pca} in the context of building generative models for functional data. fPCA on the amplitude component corresponds to an eigendecomposition of the sample covariance function across the aligned SRVFs, $\widehat{K(s,t)} = \frac{1}{k-1}\sum_{i = 1}^k ((q_i,\gamma^*_i)(s) - \hat{\mu}_q(s))((q_i,\gamma^*_i)(t) - \hat{\mu}_q(t)) = \sum_{b = 1}^\infty\hat{\lambda}_b \hat{\phi}_b(s)\hat{\phi}_b(t)$ where $\hat{\phi}_b,\ b=1,2,\dots$ are fPCs that form an orthogonal basis for the space of aligned SRVFs. With this data-driven basis, a finite representation of aligned SRVFs can be obtained through truncation as $(q_i,\gamma^*_i) \approx \hat{\mu}_q + \sum_{b = 1}^B \hat{c}_{i,b}\hat{\phi}_b$, where $\hat{c}_{i,b} = \int_0^1 ((q_i,\gamma^*_i)(s) - \hat{\mu}_q(s))\hat{\phi}_b(s) ds$; $\hat{\mathbf{c}}=\left(\hat{c}_{i,1},\ldots,\hat{c}_{i,B}\right)^\top$ is a $B$-dimensional Euclidean representation of $(q_i,\gamma^*_i)$. Then, a generative model consists of (1) drawing a random coefficient vector $\mathbf{c} \sim MVN_B(0,\text{diag}(\hat{\lambda}_1,\ldots,\hat{\lambda}_B))$, (2) constructing a random SRVF as $q_r=\hat{\mu}_q + \sum_{b = 1}^B c_b\hat{\phi}_b$, and (3) computing the corresponding function $f_r=Q^{-1}(q_r,T)$, where $T\sim N(\bar f(0),\hat{\tau}^2)$ is a random translation, and $\hat{\tau}^2$ is the sample variance of the function starting points $f_1(0),\ldots,f_k(0)$. Thus, the resulting random function $f_r$ is necessarily aligned to the mean amplitude function $\hat{\mu}_f$.

\begin{figure}[!t]
\begin{center}
\begin{tabular}{ccc}
    \includegraphics[width = 1.95 in]{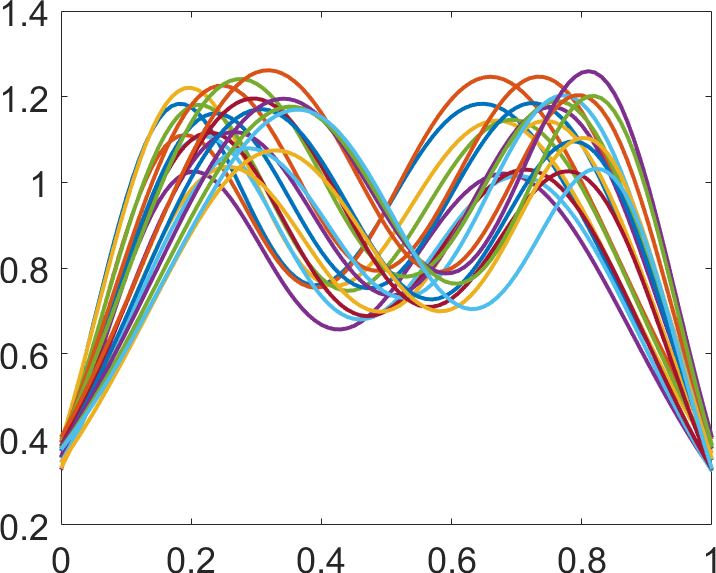} &  \includegraphics[width = 1.59 in]{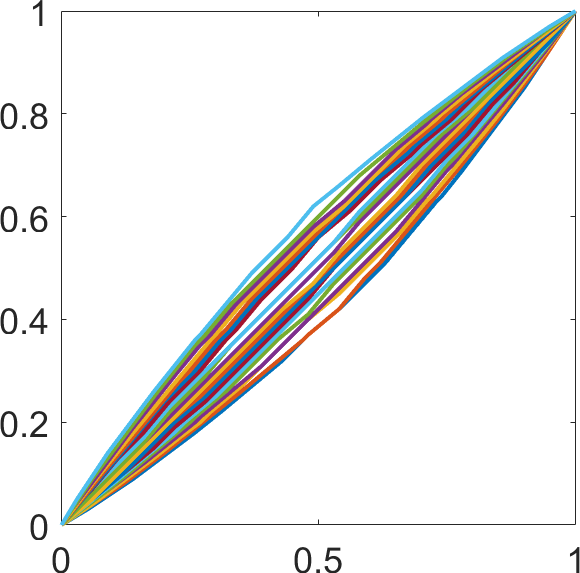}   & \includegraphics[width = 1.95in]{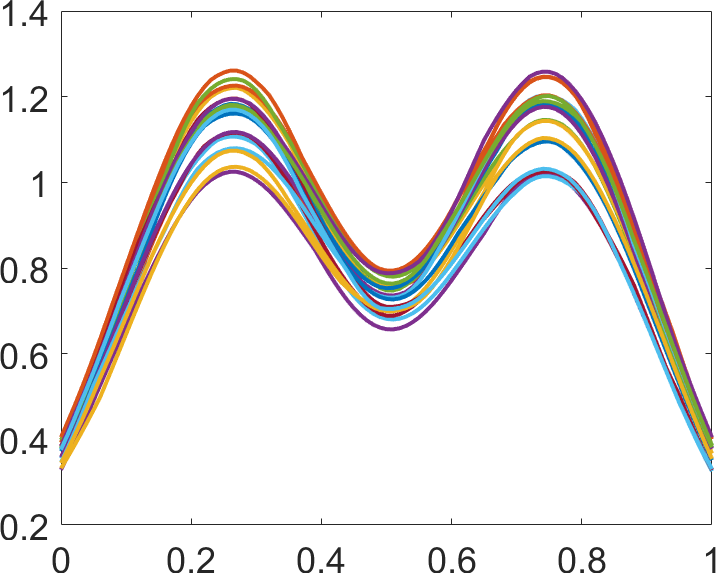}  \\
(a) & (b) & (c) \\
 \includegraphics[width = 1.95in]{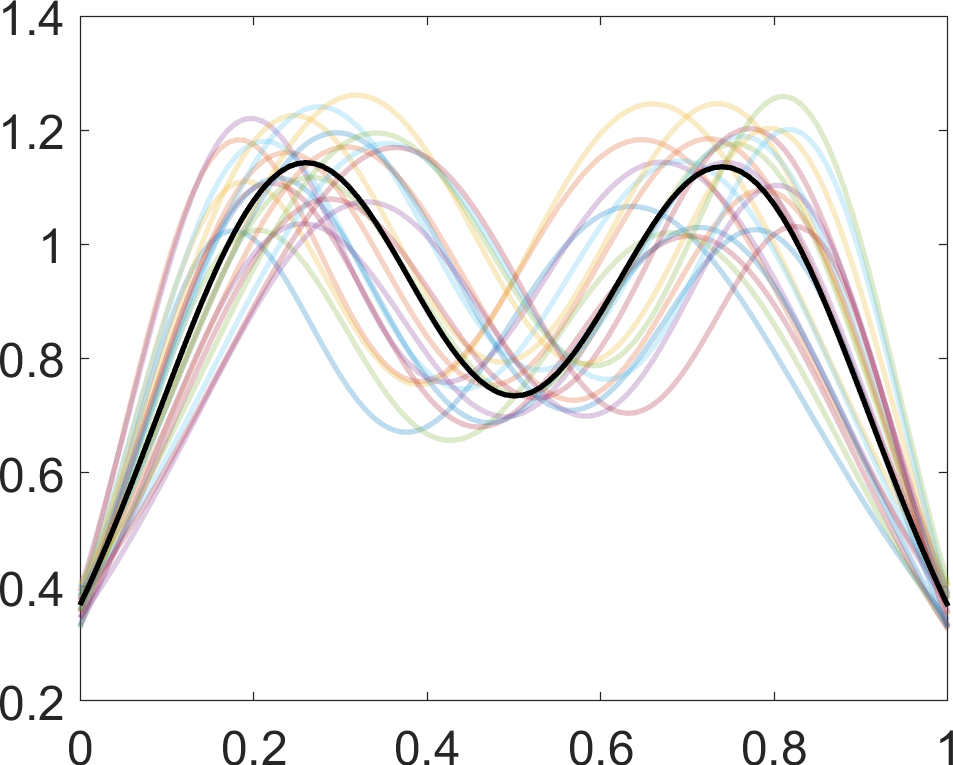} & \includegraphics[width = 1.95in]{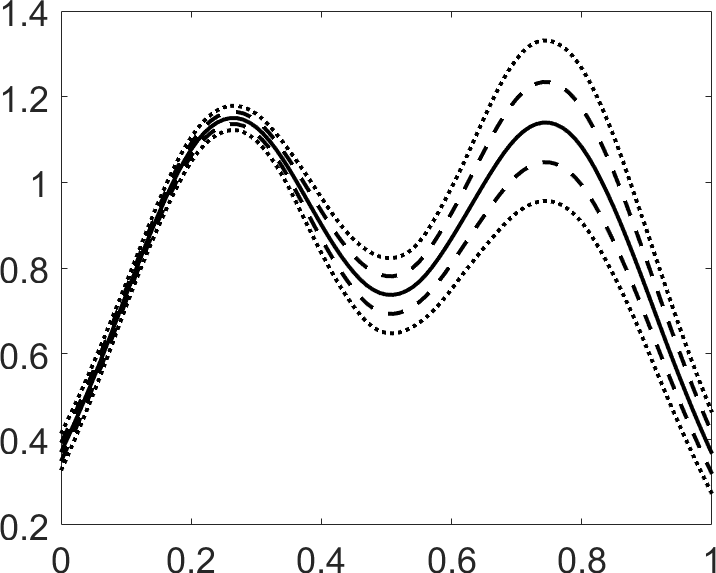}  & \includegraphics[width = 1.95in]{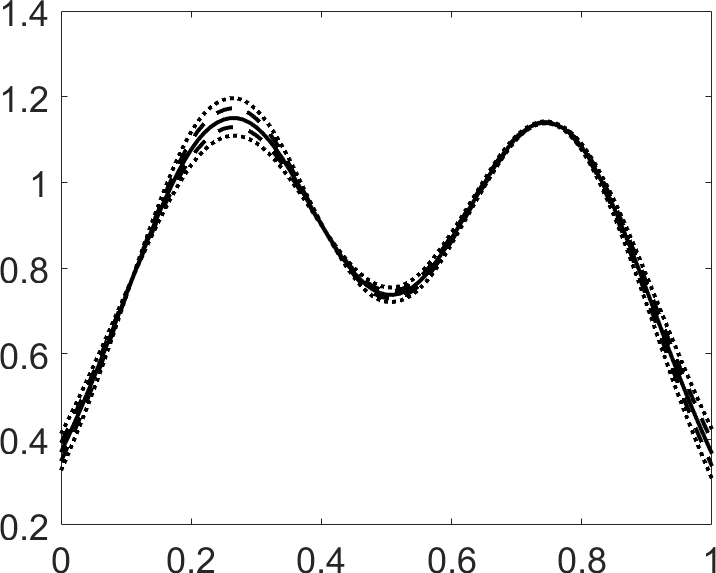} \\
    (d) & (e) & (f) \\
    \end{tabular}
    \caption{(a) Simulated, (b) phase, and (c) amplitude functions. (d) Same as (a) with mean amplitude in bold. (e) First and (f) second principal modes of amplitude variability.}
    \label{simu_summary}
    \end{center}
\end{figure}

To illustrate EFDA, we consider the sample of simulated functions shown in Figure \ref{simu_summary}(a). The phase and amplitude functions extracted via the eFR-based registration procedure are shown in panels (b) and (c), respectively, with the amplitude mean shown in bold in (d) on top of the original functions. The first two principal modes of variability for the amplitude component are given in panels (e) and (f), respectively. That is, we plot $Q^{-1}(\hat{\mu}_q + s\hat{\lambda}_b^{1/2}\hat{\phi}_b, T)$ for $b\in\{1,2\}$, $s \in \{-2,-1,0,1,2\}$; $T$ is a small translation that is included for improved display. The primary mode of variability captures differences in the height of the right peak, while the second mode describes variability in the left peak, and to a lesser extent, the relative heights of the left and right peaks. The remaining modes of variability, which are not displayed here, capture a negligible amount variability.



\section{Model-based Estimation and Registration}
\label{models}
We summarize the notation used thus far and introduce notation for discretely observed functional observations, which become key at the modeling and implementation stages. Let $\mathcal{F}$ be the set of absolutely continuous functions $f:[0,1]\to\R$, $\mathcal{Q}$ the corresponding SRVF space under the mapping $f \mapsto Q(f)=\text{sign}(\dot{f})\sqrt{|\dot{f}|}:=q$,
and $T\in\mathbb{R}$ a scalar translation. The function $Q^{-1}:\mathcal{Q}\times \R \to \mathcal{F}$ takes in an SRVF and a translation, and maps them to the corresponding point $f$ in the function space $\mathcal{F}$. A warping function is denoted by $\gamma \in \Gamma$, where $\Gamma = \{\gamma:[0,1]\rightarrow[0,1]\, | \, \gamma(0) = 0, \gamma(1) = 1, \dot{\gamma}>0\}$.

A model for functional data can be defined either on the original function space $\mathcal{F}$ or on the SRVF space $\mathcal{Q}$. While modeling phase variation through the usual warping action $f \mapsto f \circ \gamma$ is simpler on $\mathcal{F}$, the simple $\mathbb{L}^2$ geometry of $\mathcal{Q}$ makes dimension reduction through a (possibly data-driven) basis expansion possible. We choose a marriage of the two options, and assume that functional observations are generated under the model,
\begin{equation}
    \label{model}
    y_i(t)=(Q^{-1}(q_i,T_i)\circ \gamma_i)(t)+\epsilon_i(t), \quad i=1,\ldots,n,
\end{equation}
where $\epsilon_i(t)\stackrel{\text{ind}}{\sim}N(0,\sigma^2_i)$ (in most cases, one can fix all $\sigma^2_i$ to a common value), and the subscript $i$ indexes each sample element (e.g., subject). Thus, individual curves $y_i$ are pointwise perturbations of a function $Q^{-1}(q_i,T_i)$ warped by $\gamma_i$.
\emph{This allows us to specify structured prior distributions on the amplitude component based on a judicious choice of shape-defining basis functions, while specifying a prior distribution on warping functions $\Gamma$ that is naturally compatible with possible fragmentation and sparsity}. This model is characterized by subject-specific amplitude (and phase) components, in contrast to a common amplitude template as done in \cite{telesca_registration,ZSGM}. We let $\mathbf{y}=(y_1,\ldots,y_m)^\top$ and $\mathbf{t}=(t_1<\cdots<t_m)^\top$ denote a vector of noisy, discrete functional observations and the corresponding time grid on which this data was observed, respectively. Then, for a function $f$, $f(\mathbf{t}) = (f(t_1),\ldots,f(t_m))^\top$ is a vector of function evaluations on this same time grid.

\subsection{Likelihood Specification}

We now formulate error model (\ref{model}) for discretely observed functions, as is typical in practice. The vector of observations of function $i$, denoted by $\mathbf{y}_i$, is obtained from evaluations over a subject-specific grid of time points $\mathbf{t}_i = (t_{i,1}<\cdots<t_{i,m_i})^\top$ of an appropriately translated (via $T_i$) and warped (via $\gamma_i$) SRVF $q_i$ specifying the overall shape, as
\begin{eqnarray}
\label{eqn_obs}
\mathbf{y}_i = (Q^{-1}(q_i,T_i)\circ\gamma_i)(\mathbf{t}_i) + \epsilon_i(\mathbf{t}_i), \quad i = 1,\ldots,n.
\end{eqnarray}
This formulation allows general observation regimes, such as when $\mathbf{t}_i$ is sparse or fragmented. Under the normal distributional assumption on $\epsilon_i$,
the likelihood for our model, based on observation $\mathbf{y}_i$, is given by
\begin{eqnarray}
\label{eqn_likelihood}
\mathbf{y}_i| q_i, \gamma_i, T_i, \sigma_i^2 \; {\sim} \;  MVN_{m_i}\left( (Q^{-1}(q_i,T_i)\circ\gamma_i)(\mathbf{t}_i),\thinspace \sigma_i^2I_{m_i}\right), \quad i = 1,\ldots,n.
\end{eqnarray}

\subsection{Prior Models for Translation, Phase and Error Variance}
\label{prior_specification}

We first focus on model components whose probable values a-priori may be assumed relatively similar between problems. We assume that the translation parameters, $T_i,\ i=1,\ldots,n$, and error variances, $\sigma^2_i,\ i=1,\ldots,n$ are a-priori independent with $T_i\, {\sim}\, N(\mu_T,\tau^2)$ and $\sigma_i^2 \,{\sim}\,  \text{Inverse-Gamma}(\alpha_\sigma,\beta_\sigma)$. The choice of hyperparameters $\tau^2,\ \alpha_\sigma$ and $\beta_\sigma$ is largely problem-specific. In all applications, we use $\mu_T=0$.

Defining a prior process on warping functions is more challenging due to its restricted functional form. Common approaches in the literature model phase functions using basis expansions with constrained coefficients \cite{fu_registration,telesca_registration} or directly via the Riemannian geometry of the group of warping functions \cite{lu_registration}. In contrast, we use a piecewise linear process proposed in Bharath and Kurtek \cite{bharath_phase}, consisting of random phase increments $p(\gamma_i) = (\gamma_i(t_1),\ldots,\gamma_i(t_j) - \gamma_i(t_{j-1}),\ldots,1 - \gamma_i(t_{m_\gamma}))^\top$ over $m_\gamma$ successive time points $t_1<\cdots<t_{m_\gamma}$ on the input domain $[0,1]$. The partition size is a user-specified value $m_\gamma \leq \underset{i}{\text{min } }m_i$, whose magnitude provides a trade-off between model flexibility and computation time.
Each finite-dimensional vector of phase increments follows a Dirichlet distribution, $p(\gamma_i) \overset{\text{iid}}{\sim} Dirichlet(\theta_{\gamma}  p(U^*(0,1)))$, where $U^*(0,1)$ is a vector of order statistics from a $\text{Uniform}(0,1)$ random sample of size $m_\gamma$, and $\theta_\gamma$ acts as a precision parameter. This finite-dimensional model specification is computationally convenient; furthermore, Bharath and Kurtek \cite{bharath_phase} show that it has desirable asymptotic properties by relating it to a stochastic process with (measurably) increasing sample path as the time increments become arbitrarily small.

In this prior specification, the choice of the order statistics dictates where the prior process is centered. For example, uniform order statistics result in a prior distribution over $\Gamma$ centered at the identity warping $\gamma_id$; order statistics from an arbitrary distribution $G$ on $[0,1]$ would result in the distribution being centered at $G^{-1}$. We choose the uniform order statistics to regularize the prior model toward the identity warping. We evaluate the prior approximately by assigning the random order statistics to a uniform spacing of the domain grid, since the successive spacings in both cases are essentially of order $O(1/m_\gamma)$. The precision hyperparameter $\theta_\gamma$ controls the spread of the prior over $\Gamma$: a small value results in a diffuse prior, whereas a large value concentrates the prior around its mean warping. The choice of $\theta_\gamma$ should in part depend on the number of discrete time observations in the data (observational regime), i.e., when the functions are densely sampled $\theta_\gamma$ can be small, but when faced with sparse data, a considerable amount of regularization is required to identify subject-specific model components.


\subsection{Shape-informed Prior Models for Amplitude}
\label{amplitude_models}

Besides allowing direct interpretation of phase and amplitude uncertainty, modeling underlying functions in model (\ref{eqn_obs}) via the EFDA framework allow us to define the amplitude model on the mathematically convenient SRVF representation space. Choice of the prior model over the registered SRVFs, representing amplitude, is both problem-specific and has a potentially strong impact on posterior inference under general observation regimes.
We propose two semi-automatic approaches to model prior information about amplitude in different data collection scenarios. The first type is appropriate when, along with data that is sparsely observed or fragmented, we also have access to functional observations of different subjects from the same population that are neither fragmented nor sparse; we refer to this data as training and define an empirical prior model based on statistics computed from this data. The second type of prior is appropriate when information is available about the maximum number and relative locations of local extrema that underlie the functional observations. Importantly in practice, this prior model does not require the location of the extrema on the domain. We will also explain how this second scenario has connections to landmark registration \cite{kneip_landmark}.

\subsubsection{Empirical Amplitude Prior}

\begin{figure}[!t]
\begin{center}
\begin{tabular}{ccc}

  \includegraphics[width = 1.5 in]{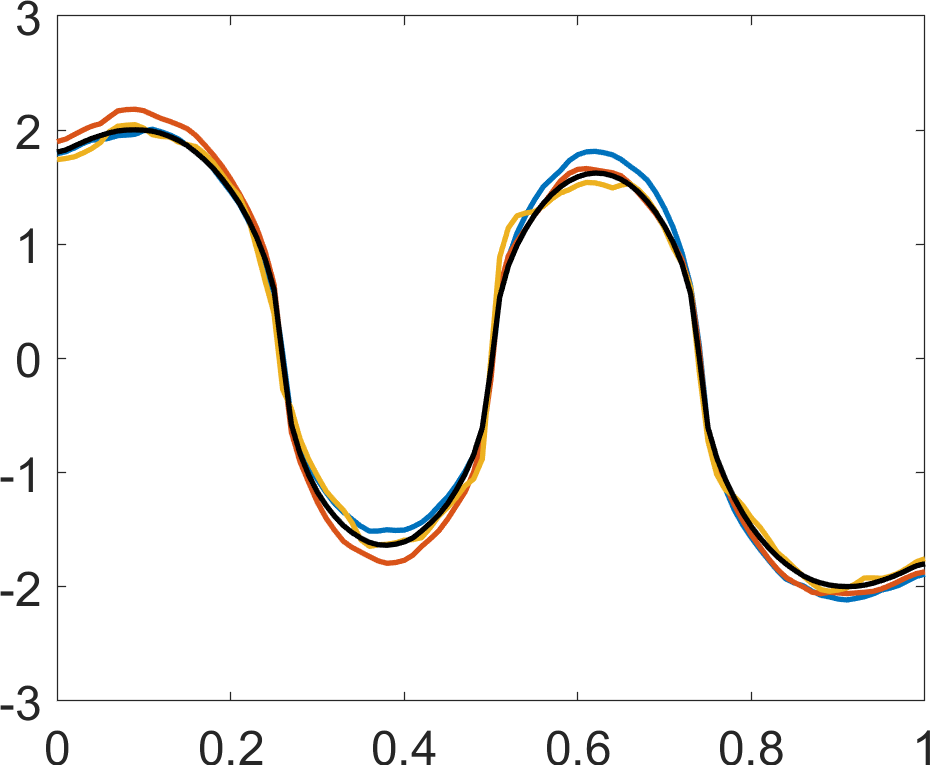}  & \includegraphics[width = 1.5 in]{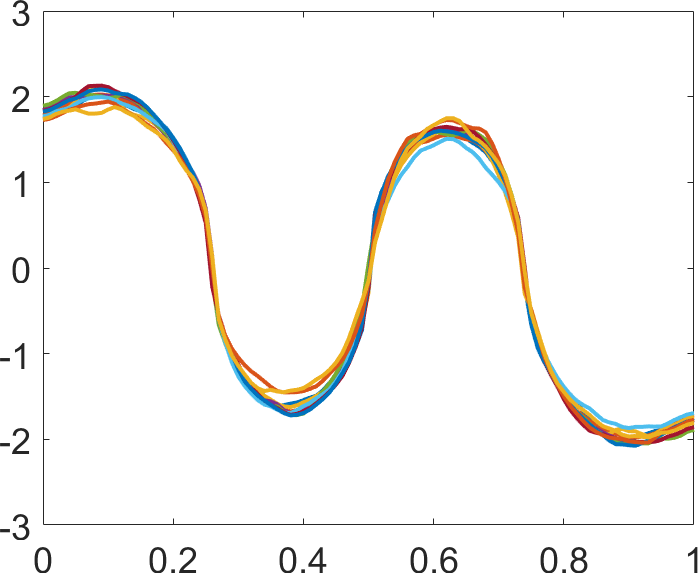} & \includegraphics[width = 1.61 in]{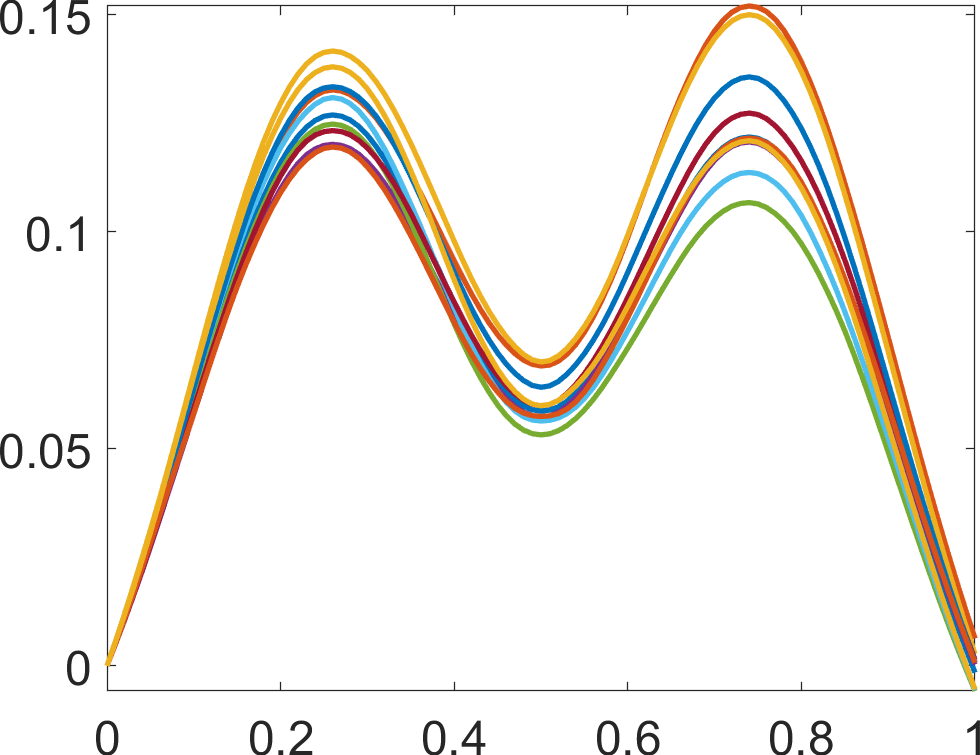} \\
    (a) & (b) & (c) \
    \end{tabular}
    \caption{(a) Sample mean SRVF (black) and fPCA basis elements computed using the training data in Figure \ref{simu_summary}. (b) SRVF draws, and (c) corresponding amplitude functions transformed under $Q^{-1}$, generated from the proposed empirical amplitude prior model.}
    \label{PCA_prior}
    \end{center}
\end{figure}

When densely-observed training data $f_1,\ldots, f_k$ is available, we can construct a shape-informed prior for the amplitude of $n$ new, partially observed functions on the subspace spanned by the empirical basis constructed from the amplitudes of training data SRVFs $q_1,\ldots,q_k$. Elastic fPCA is carried out by first jointly performing registration and computation of the Karcher mean $\hat{\mu}_q$, and then decomposing the sample covariance of the amplitude components of the training data to obtain eigenfunctions $\{\hat{\phi}_b,\ b=1,\ldots,B\}$ and corresponding eigenvalues $\hat \lambda_b$,\ $b=1,\ldots,B$. The basis functions $\hat{\phi}_b$ represent amplitude variation about the Karcher mean $\hat{\mu}_q$ so that the SRVFs $q_i,\ i=1,\ldots, n$ in model (\ref{model}) can be represented as
\begin{eqnarray}
\label{eqn_pca}
q_i = \hat{\mu}_q + \sum_{i=b}^B c_{i,b}\hat{\phi}_b,\quad i = 1\ldots,n.
\end{eqnarray}
Thus, a prior process over the amplitude component $q_i$ can be defined through a prior distribution over the coefficient vector $\mathbf{c}_i$, such as $\mathbf{c}_i \sim  MVN_B(0_B,\text{diag}(\hat{\lambda}_1,\ldots,\hat{\lambda}_B))$.  
This prior specification provides a data-driven way of imposing structure on the amplitude model based on the training data. Note that the training observations must be representative of the important features of amplitudes in the population, and their number must be sufficient to be able to estimate these. The proposed prior model is illustrated in Figures \ref{simu_summary} and \ref{PCA_prior} for a simulated training dataset. Figure \ref{PCA_prior}(a) shows the sample mean SRVF (black) and fPCA basis computed using the amplitude functions in Figure \ref{simu_summary}(c). SRVF draws from the prior, and corresponding amplitude functions obtained by the transformation $Q^{-1}$, are shown in Figure \ref{PCA_prior}(b)-(c), respectively. We illustrate visually that these random functions are registered to the amplitude functions in the training data, as required.

\subsubsection{Shape-restricted Amplitude Prior}

When reliable information about the number and relative ordering of extrema of underlying functions is available, a prior on the amplitude $q_i$ can be specified by choosing a set of basis functions that reflects this information. In contrast to the empirical amplitude prior, such a prior does not require training data. We assume the following form for the SRVFs $q_i$:
\begin{eqnarray}
\label{eqn_restrictedspline}
q_i = \sum_{i=b}^Bc_{i,b}U^*_b,\quad i = 1\ldots,n. \end{eqnarray}
The basis functions are defined as
$U^*_b(t) = M (\prod_{h = 1}^H(t - \alpha_h))U_b(t)$, $b=1,\ldots,B$, where $U_b$ are B-spline basis functions. This basis system is based on a modification of the shape-restricted B-splines developed by Wheeler et al. \cite{wheeler_spline}. These bases relate to the derivative of the function since amplitudes are defined on the SRVF space rather than the original function space. This basis system forces the SRVFs to be exactly zero at the change point locations $\alpha_1,\ldots,\alpha_H$, corresponding to extreme values at these locations in the corresponding amplitude functions. The constant $M \in \{-1,1\}$ is application-specific and defines the order of extreme values.

Our use of this basis system differs from \cite{wheeler_spline} as our model accounts for phase variability explicitly, i.e., we treat the change points as constants that determine the common locations of extreme values of the underlying amplitude functions, unlike the original work where the change points are also inferred. We use a diffuse exponential prior model on the coefficient vector $\mathbf{c_i}$ to ensure that each basis coefficient is positive; this, in turn, guarantees that all of the amplitude functions have the same ordering of extrema. For this prior specification, the size of the discretization grid for the phase functions, $m_\gamma$, cannot be set arbitrarily; to ensure identifiability, we set $m_\gamma = H$.

The shape-restricted amplitude prior model has a clear connection to landmark-based registration \cite{kneip_landmark} in which one must first identify a set of common landmarks on each function in a dataset. The landmarks either correspond to mathematical features of the data, e.g., extrema, or application-specific, interpretable features. Given a set of landmarks on each function, the functions are registered to each other via a piecewise linear warping that aligns the landmark locations exactly. If only extrema are considered as landmarks, then the change points in the shape-restricted amplitude prior act as domain locations for landmark alignment. In general, selecting appropriate landmarks can be challenging, especially when the number of functions and/or landmarks is large. The process of selecting non-mathematical landmarks can also be highly subjective. Recently, Strait et al. \cite{strait_landmarks} developed an automated approach for mathematical landmark selection that alleviates the aforementioned issues.

\begin{figure}[!t]
\begin{center}
\begin{tabular}{ccc}
  \includegraphics[width = 1.65 in]{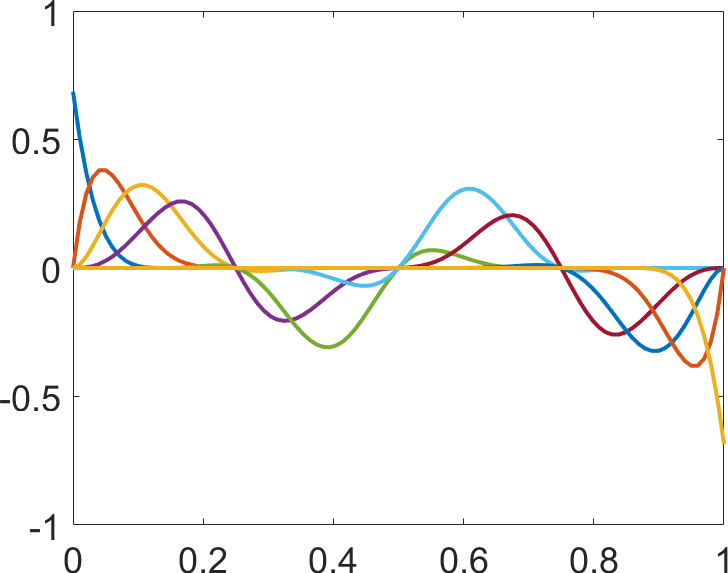}  & \includegraphics[width = 1.69 in]{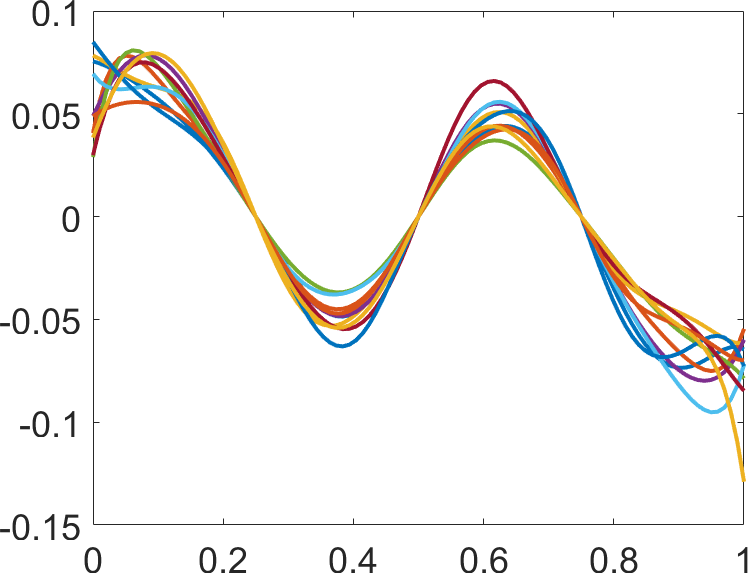} & \includegraphics[width = 1.59 in]{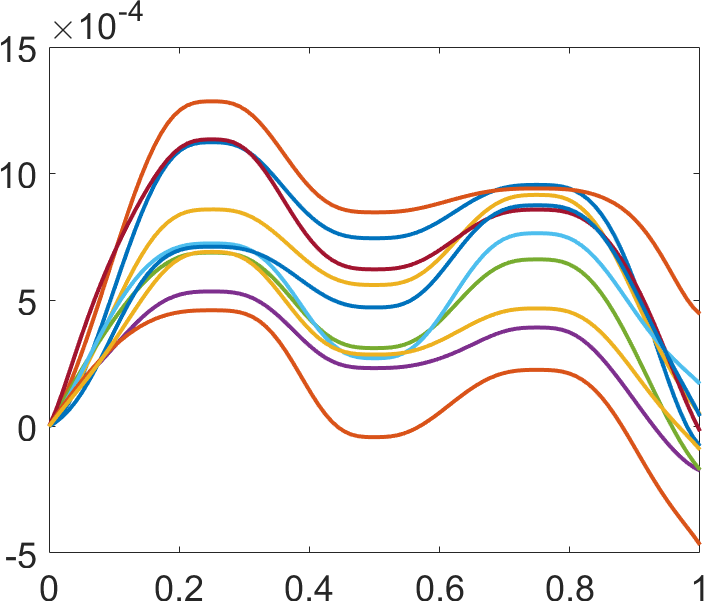} \\
    (a) & (b) & (c) \\
    \end{tabular}
    \caption{(a) Shape-restricted spline basis. (b) SRVF draws, and (c) corresponding amplitude functions, generated from the proposed shape-restricted amplitude prior model with $H = 3,\ \mathbf{\alpha}=(0.25, 0.5,0.75)^\top,\ M = -1$.}
    \label{rs_basis}
    \end{center}
\end{figure}

Figure \ref{rs_basis} illustrates the shape-restricted amplitude prior and the structure it enforces on the amplitude component in our observation model. Panel (a) depicts the basis system that is used to represent the SRVFs. Panels (b) and (c) show several prior draws of SRVFs and corresponding amplitude functions, respectively. The basis system is constrained to take zero values at the change point locations. Combined with the appropriate restrictions on $M$ and the basis coefficients, this pattern propagates to the generated SRVFs. Zeros of SRVFs map to extrema of amplitude functions; the relative heights of the extrema are flexible.

Both methods proposed here for informing the amplitude model consist of constructing a set of meaningful basis elements, either empirically based on training data or from prior shape information concerning the number and relative location of extrema. The shape-restricted amplitude prior relies on the practitioner to specify the number and pattern of extreme values, while the empirical amplitude prior achieves this automatically, but requires training data. Another difference is that functions in the amplitude space spanned by the empirical basis are aligned with respect to the eFR metric, while elements of the shape-restricted amplitude space are only registered based on the extrema. In this sense, the empirical basis is far more informative than the shape-restricted basis as illustrated in Figures \ref{PCA_prior} and \ref{rs_basis}.

\section{Simulations and Real Data Applications}

In this section, we discuss several simulated and real data examples which illustrate the performance of the proposed framework. We compare our methodology to PACE, implemented via a publicly available software package \cite{yao_pace,tang_registration}. The PACE framework is the most appropriate for comparison to our approach as it is able to accommodate sparsity, pointwise noise, and phase and amplitude variability in functional data. In brief, the PACE approach uses three steps: (1) estimation of population parameters, such as the mean and principal components, based on noisy and sparse observations, (2) estimation of fitted functional observations on a common domain grid based on the parameter estimates from (1) \cite{yao_pace}, and (3) registration of the estimated functions via a penalized $\mathbb{L}^2$ metric criterion \cite{tang_registration}. There are two versions of the procedure, one for observations recorded over a common grid of domain points and one for observations recorded on different grids. When observations are recorded on a common grid, the PACE procedure begins with steps (1)-(3) (step (3) results in the final phase functions). Then, the estimate from step (2) is disregarded; instead, the phase functions are applied to the original noisy observations followed again by steps (1) and (2) to estimate the final amplitude functions. In this setting, a single estimate of the original observations is generated through composition of the final phase and amplitude estimates (termed PACE in results). On the other hand, when observations are recorded on different domain grids, following a first iteration through steps (1)-(3) (step (3) here results in a first estimate of amplitude functions and the final phase functions), steps (1) and (2) are applied again to the registered estimated functions to extract a second estimate of amplitude. Thus, one can actually obtain two different estimates: the first through composition of the final phase functions and the first estimate of amplitude functions (termed PACE in results), and the second through composition of the final phase functions and the second estimate of amplitude functions (termed WPACE in results). All tuning parameters are set by default in the package. In the simulation examples, we consider three estimators of a sparsely-observed or fragmented function $f$ from a posterior MCMC sample $\{q^{[j]},\  j=1,\ldots,N\}$ after burn-in:
\begin{enumerate}
    \item The plug-in estimator $\hat f_{\text{plug-in}} = \left[\frac{1}{N}\sum_{j=1}^N Q^{-1}(q^{[j]},T^{[j]})\right]\circ\left[\frac{1}{N}\sum_{j=1}^N\gamma^{[j]}\right]$.
    \item The maximum a-posteriori (MAP) estimator $\hat f_{\text{MAP}} = Q^{-1}(q^{[p]},T^{[p]})\circ\gamma^{[p]}$ where $p$ is the index with the largest unnormalized log-posterior density.
    \item The pointwise estimator $\hat f_{\text{pointwise}} = \frac{1}{N}\sum_{j=1}^N \left[Q^{-1}(q^{[j]},T^{[j]})\circ\gamma^{[j]}\right]$.
\end{enumerate}
The performance of each estimator is compared with the PACE and WPACE estimators.

\label{examples}
\subsection{Bayesian Model with Empirical Amplitude Prior}

We begin the results section with a few examples that use the empirical amplitude prior in the proposed framework to fit fragmented and sparse functions.

\subsubsection{Simulated Example 1: Fragmented Simulated Functions}

\begin{figure}[!t]
\begin{center}
    \begin{tabular}{ccc}
  \includegraphics[width = 1.95 in]{figures_whitespace/simu_trainingf.png} &  \includegraphics[width = 1.95 in]{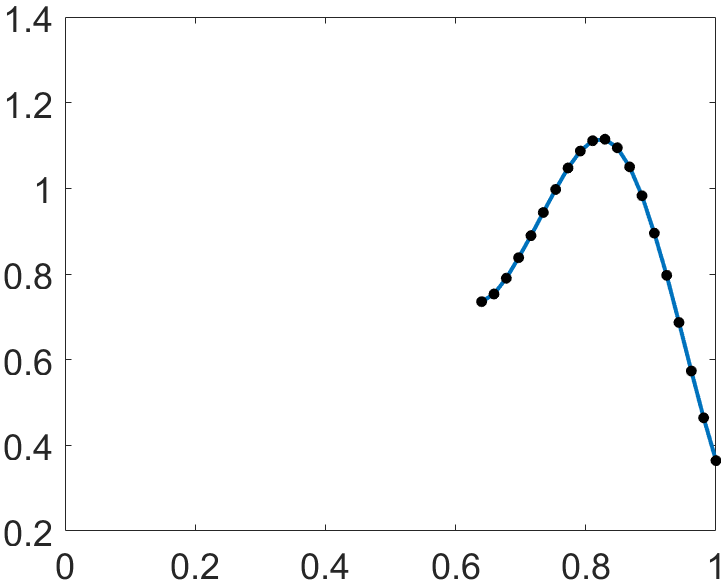}& \includegraphics[width = 1.95 in]{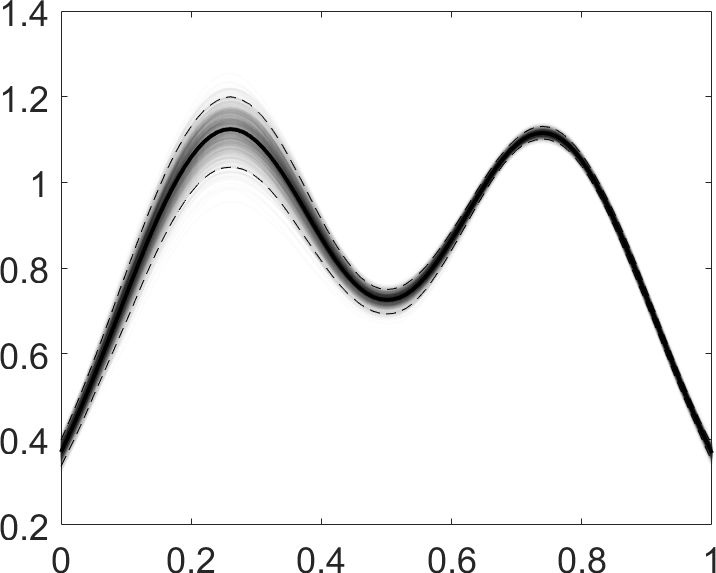}\\
    (a) & (b) & (c) \\
	 \includegraphics[width = 1.59 in]{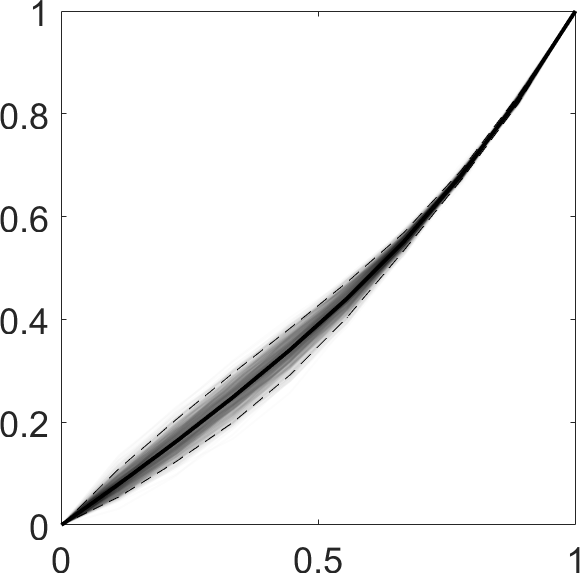}  & \includegraphics[width = 1.96 in]{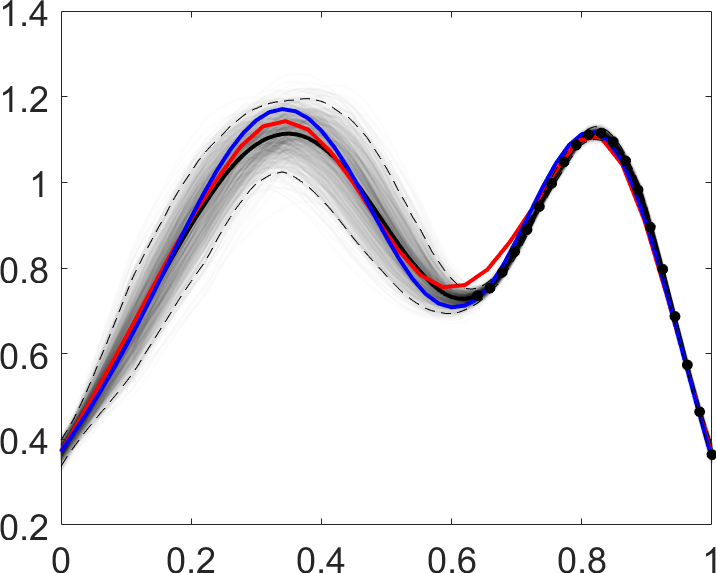}&\includegraphics[width = 1.99in]{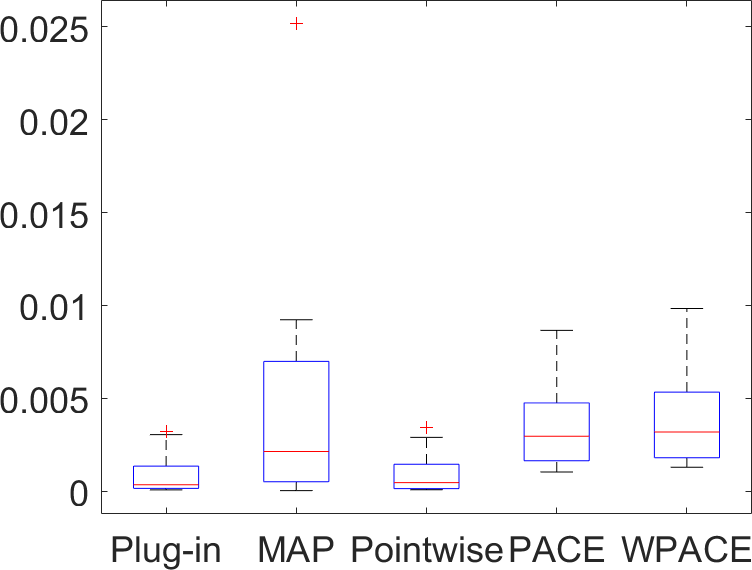} \\
    (d) & (e) & (f) \\
    \end{tabular}
    \caption{(a) Simulated training data. (b) Fragmented simulated function with observations shown as black points. Posterior draws (transparent), pointwise mean (solid black) and $95\%$ credible interval (dashed) for the (c) amplitude and (d) phase components. (e) Composition of amplitude and phase, and PACE (red) and WPACE (blue) estimates. (f) Boxplots of $\mathbb{L}^2$ distances between the true function and estimated function. Each boxplot corresponds to a different approach: (1) plug-in, (2) MAP, and (3) pointwise estimators based on posterior samples from the proposed Bayesian model, and (4) PACE and (5) WPACE.}
    \label{fragmented_fit}
    \end{center}
\end{figure}

Figure \ref{fragmented_fit}(a)-(b) shows simulated training data and a fragmented functional observation (black points) that we wish to infer, respectively. Within our unified framework, we employ an empirical amplitude prior with $\theta_\gamma = .1,\ m_\gamma = 8$ and $B = 8$, to infer the full underlying function.  Figure \ref{fragmented_fit} shows posterior marginal samples over the amplitude (c) and phase (d), and their composition (e); the solid black line in each panel represents the pointwise posterior mean. This display allows us to assess uncertainty in the different components underlying the estimated function. Additionally, in (e), we show the PACE and WPACE estimates in red and blue, respectively, for comparison. All three approaches provide reasonable estimates with a peak along the missing portion of the function; this is in agreement with the training data, and the estimated peaks across the three different methods are in roughly similar regions of the domain. 

A major advantage of the proposed approach over PACE and WPACE is its ability to assess uncertainty in the estimated function. As evidenced by the credible bands, uncertainty is smaller along the portion of the function where data was observed than the portion that was not observed, both in terms of the height and location of the missing peak.

To provide a quantitative comparison of our method with PACE/WPACE, we simulated 100 missing portions of different functions from the complete data in Figure \ref{fragmented_fit}(a). In each simulation, one function from the complete dataset was selected at random and fragmented from $t=0$ up to a random point on $[0,1]$ drawn from a $\text{Beta}(25,25)$ distribution; the remaining functions were treated as training data. The observation set in each simulation consisted of 20 evenly-spaced points along the non-fragmented part of the randomly selected function. As our inferential approach is based on the posterior distribution over unknown model components, we consider the three estimators $\hat f_{\text{plug-in}}$, $\hat f_{\text{MAP}}$ and $\hat f_{\text{pointwise}}$ described above. We compare our performance to PACE/WPACE by computing the $\mathbb{L}^2$ distance between the true function and the estimate. Boxplots of these distances for each method are shown in Figure \ref{fragmented_fit}(f). The MAP estimator appears to be comparable in performance to PACE and WPACE. However, both the plug-in and pointwise estimators outperform PACE in $96\%$ and WPACE in $97\%$ of the simulations. We remark that the behavior of the MAP estimate is highly sensitive to inference of the phase function. In the fragmented region, where no values of the function are observed, the posterior distribution is driven largely by the diffuse prior. Consequently, the MAP estimate generally fits the data very well in the non-fragmented region, but exhibits random warping in the fragmented region; this results in a high level of misalignment between the estimated and true functions in the fragmented region, which is greatly penalized by the $\mathbb{L}^2$ distance. That said, the estimated \emph{amplitude} portion of the MAP samples is very accurate and reflects the shape of the true function in terms of the heights of the peaks and valley.

\subsubsection{Simulated Example 2: Sparse ECG Signals}

\begin{figure}[!t]
\begin{center}
\begin{tabular}{cc}
 \includegraphics[width = 2.5 in]{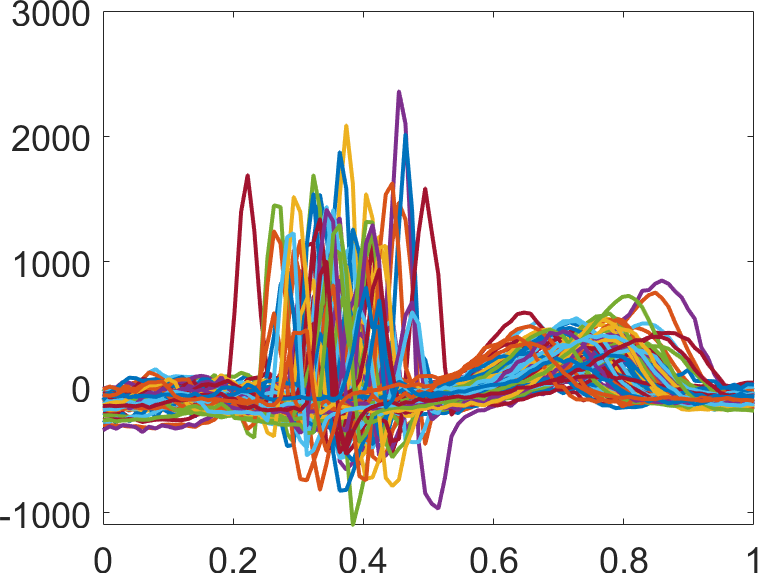} & \includegraphics[width =2.5 in]{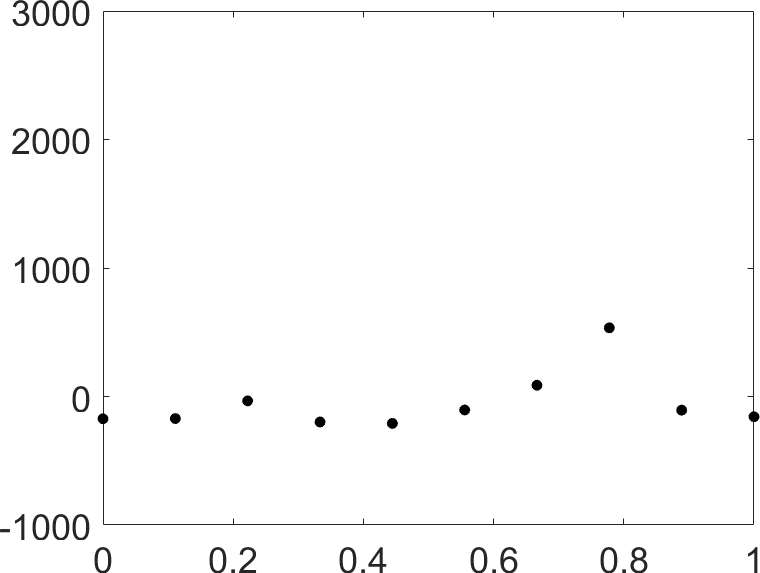}  \\
  (a) & (b) \\
\end{tabular}
\begin{tabular}{ccc}
    \includegraphics[width = 1.95 in]{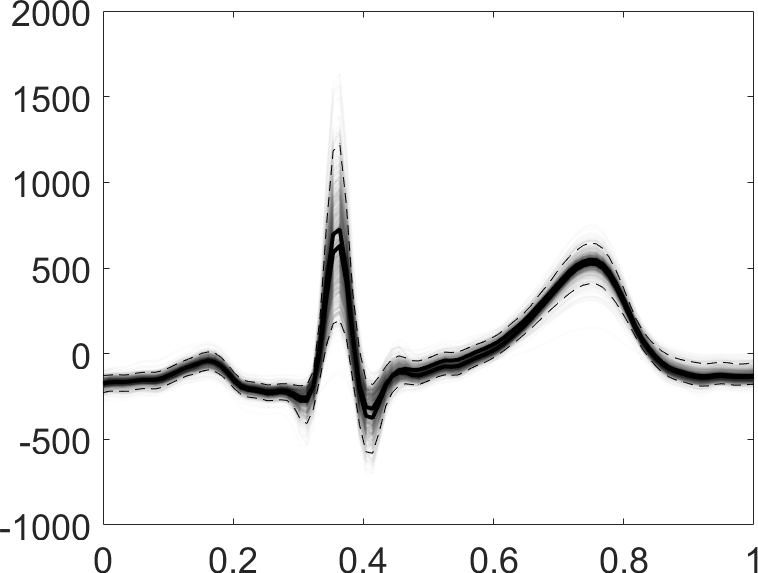} & \includegraphics[width =1.49 in]{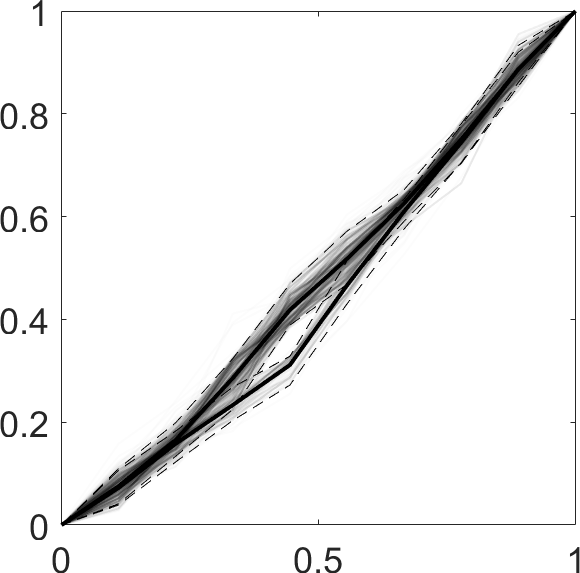} &  \includegraphics[width = 1.95 in]{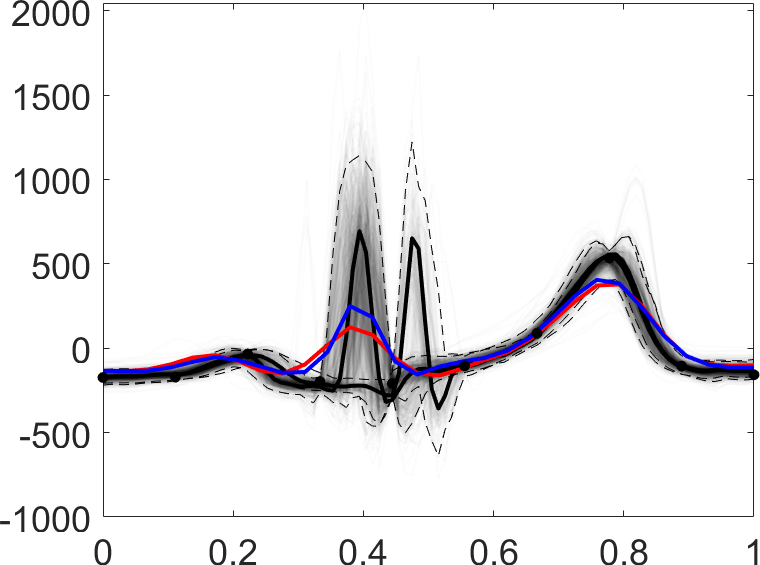}\\
    (c) & (d) & (e) \\
    \end{tabular}
    \caption{(a) ECG training data. (b) Sparsely observed ECG signal. Posterior draws (transparent), pointwise mean (solid black) and $95\%$ credible interval (dashed) for the (c) amplitude and (d) phase components. (e) Composition of amplitude and phase, and PACE (red) and WPACE (blue) estimates.}
    \label{sparse_fit}
    \end{center}
\end{figure}

The electrocardiogram (ECG) is an important diagnostic tool for many conditions including myocardial infarction. It records fluctuations in electrical potential of the heart muscle on the body surface. Often, one studies the shape of PQRST complexes extracted from a long ECG signal, which can be associated with abnormal heart function \cite{kurtek_ecg}. The letters PQRST refer to the first peak (P wave), the shallow, deep valley followed by the sharp second peak and another valley (QRS complex), and finally the third peak (T wave). In this simulation example, we study the performance of the proposed Bayesian model in the context of PQRST complex estimation from sparse observations. Figure \ref{sparse_fit} shows a set of training signals in (a) as well as a sparse set of ten evenly-spaced data points extracted from a known PQRST complex (not part of the training data) in (b); such a setup allows us to again assess the performance of our approach qualitatively and quantitatively. We employ an empirical amplitude prior under the settings $\theta_\gamma = 100,\ m_\gamma = 8$ and $B = 10$. We display marginal posterior samples in Figure \ref{sparse_fit} for amplitude (c), phase (d) and their composition (e) describing posterior uncertainty in the unknown PQRST complex function. It turns out that the posterior distribution in this case is bimodal, with two modes formed by the phase functions. Thus, we display modewise summaries for the phase sample in panel (d) and the composition of amplitude and phase in (e). Again, the means are shown in bold black and the $95\%$ confidence bands as dashed lines. The two modes correspond to two plausible locations of the QRS complex given the data. Indeed, none of the sparse observations cover the sharp R peak making its location difficult to predict. Importantly, the structure of the estimated PQRST complex based on each mode of the posterior distribution is valid. In contrast, the QRS complex in the PACE and WPACE estimates is highly distorted.

\begin{figure}[!t]
\begin{center}
\begin{tabular}{cccc}
 \includegraphics[width = 1.38 in]{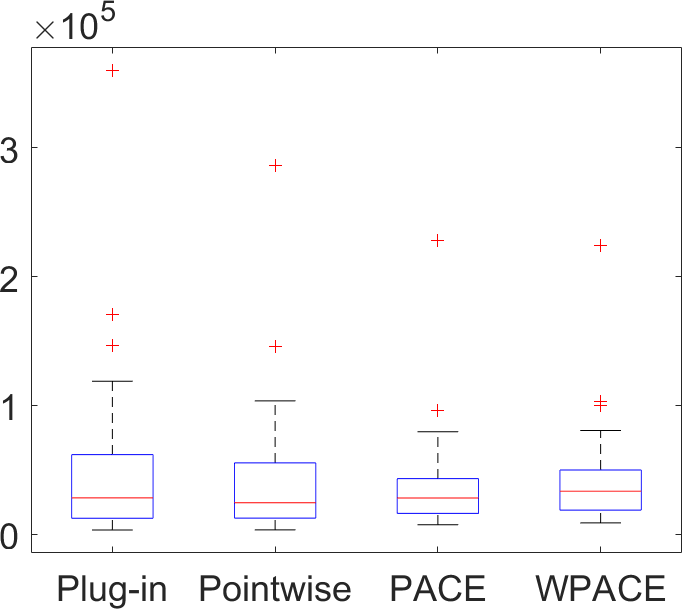} &
 \includegraphics[width = 1.38 in]{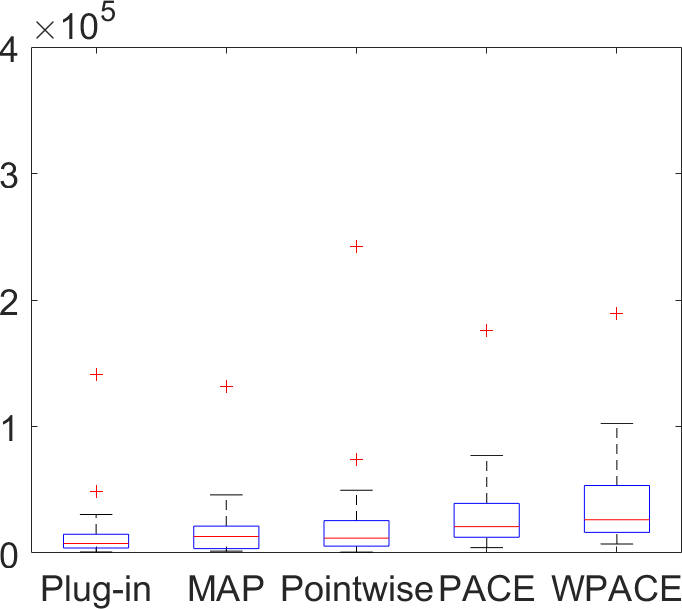}&
 \includegraphics[width = 1.5 in]{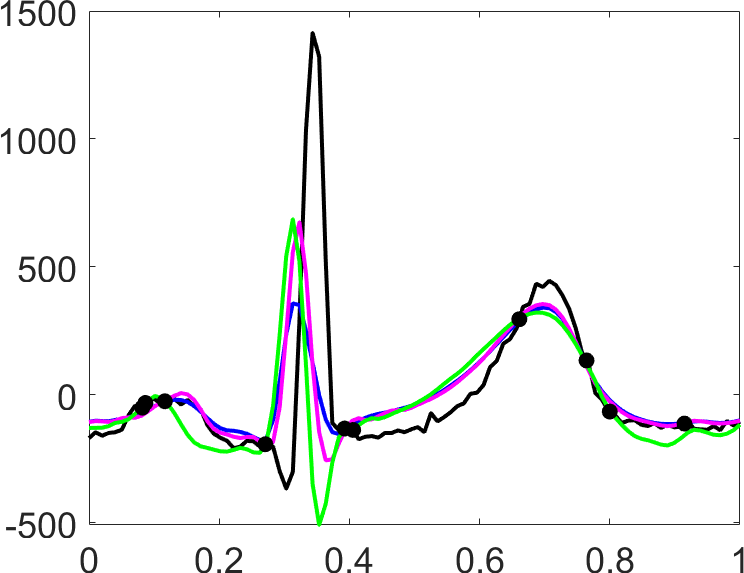} & \includegraphics[width =1.5 in]{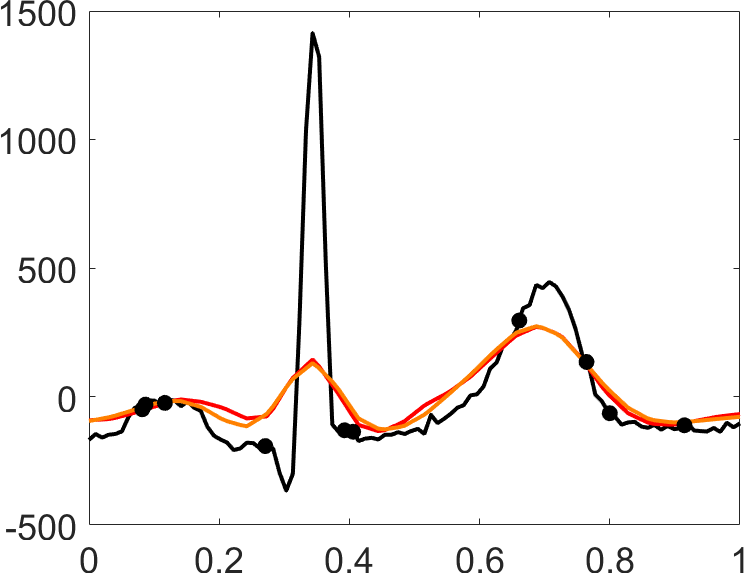}\\
  (a) & (b) & (c) & (d)\\
    \end{tabular}
    \caption{(a) Boxplots of $\mathbb{L}^2$ distances between the true function and estimated function. Each boxplot corresponds to a different approach: (1) plug-in, (2) MAP, and (3) pointwise estimators based on posterior samples from the proposed Bayesian model, and (4) PACE and (5) WPACE. (b) Same as (a) but for the amplitude component only. (c) The true function (black) and corresponding sparse observations (black points), and plug-in (magenta), MAP (green) and pointwise estimates (blue). (d) Same as (c) but with PACE (red) and WPACE (orange) estimates.}
    \label{sparse_MSE}
    \end{center}
\end{figure}

As with the preceding example, we numerically assess estimation performance of all methods. In this simulation, for each of 100 iterations, we select a PQRST complex from the training data at random, and artificially subsample it to ten observations chosen independently and uniformly along the domain; the remaining PQRST complexes are treated as training data to generate the empirical prior for the amplitude. We consider the same three posterior estimates as in the previous section and compare them to the PACE and WPACE estimates. Boxplots of the $\mathbb{L}^2$ distance from the true function to the five different estimates are shown in Figure \ref{sparse_MSE}(a). It appears that all methods show comparable performance.

Consensus amongst the different estimators occurs mainly due to the fact that the $\mathbb{L}^2$ distance criterion used here greatly penalizes misalignment between the true and estimated functions. Consider the example visualized in Figure \ref{sparse_MSE}(c)-(d). The three estimates in (c) are based on the proposed model, while the two estimates in (d) correspond to PACE and WPACE. The estimated functions in panel (c) are clearly better at capturing the shape of the PQRST complex. Unfortunately, the estimated phase results in a slight misalignment of the very pronounced R peak, which carries a significant penalty based on the $\mathbb{L}^2$ distance. This misalignment in the estimate is due to a lack of observations along this important feature of the PQRST complex. On the other hand, the PACE and WPACE estimates in panel (d) are not at all successful at capturing the true shape of the PQRST complex. To confirm this behavior, we additionally display the $\mathbb{L}^2$ distances between the amplitude components of the estimated PQRST complexes and the true amplitude component. This is done by first optimally aligning the estimates to the true PQRST complex using the eFR metric. The corresponding boxplots are displayed in Figure \ref{sparse_MSE}(b). It is clear that, with respect to this measure, the proposed model recovers amplitude features much better than PACE or WPACE. In fact, the plug-in estimate performs better than the PACE and WPACE estimates in $95\%$ and $98\%$ of the 100 iterations, respectively. Similarly, the MAP (pointwise) estimates perform better than the PACE and WPACE estimates in $85\%$ ($83\%$) and $97\%$ ($97\%$) of the 100 iterations, respectively.

\subsubsection{Real Data Example: FA from DT-MRI}



\begin{figure}[!t]
\begin{center}
\begin{tabular}{ccc}
    \includegraphics[width = 1.95 in]{figures_whitespace/FA_training.png} & \includegraphics[width = 1.95 in]{figures_whitespace/FA_missing.png} & \includegraphics[width = 1.99 in]{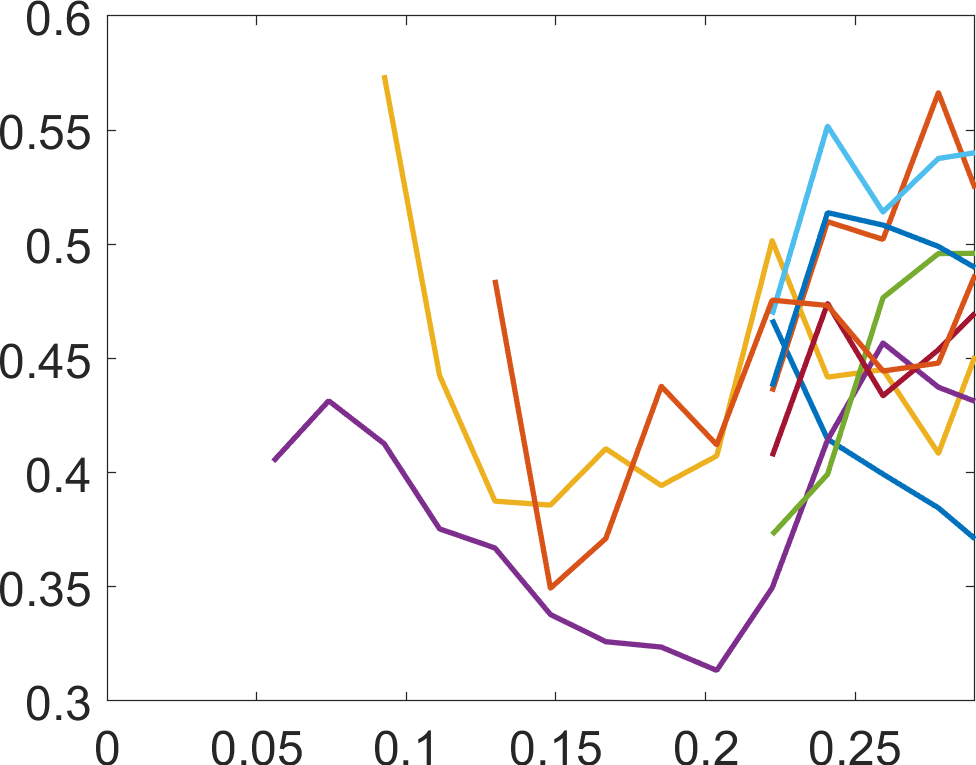}\\
    (a) & (b)& (c) \\
    \includegraphics[width = 1.95 in]{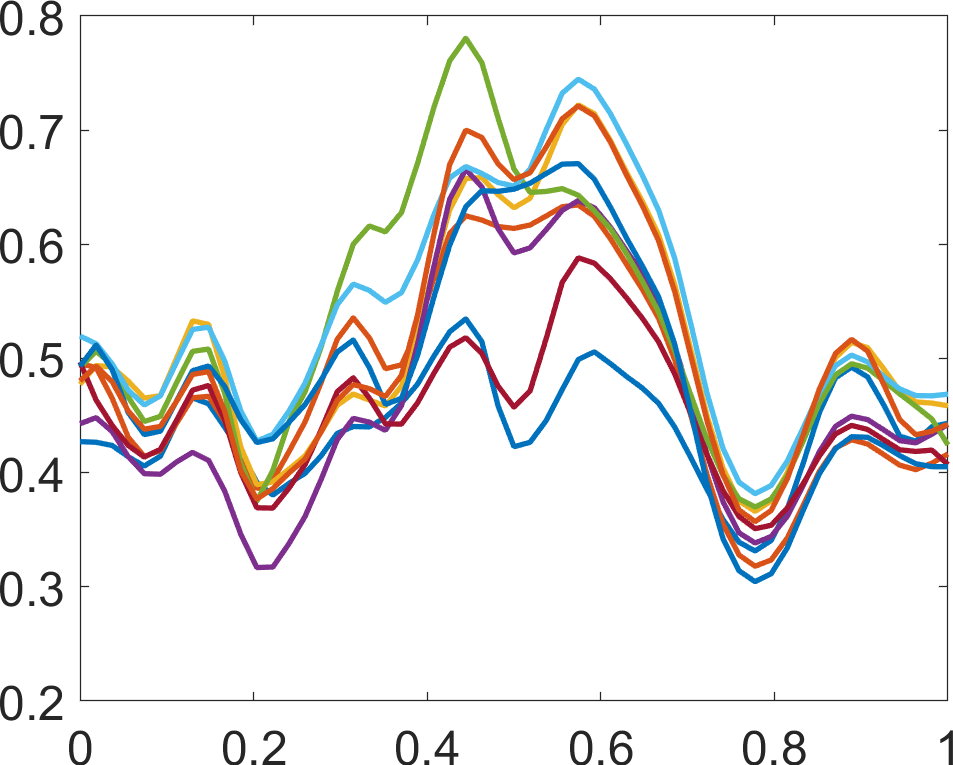}   & \includegraphics[width = 1.58 in]{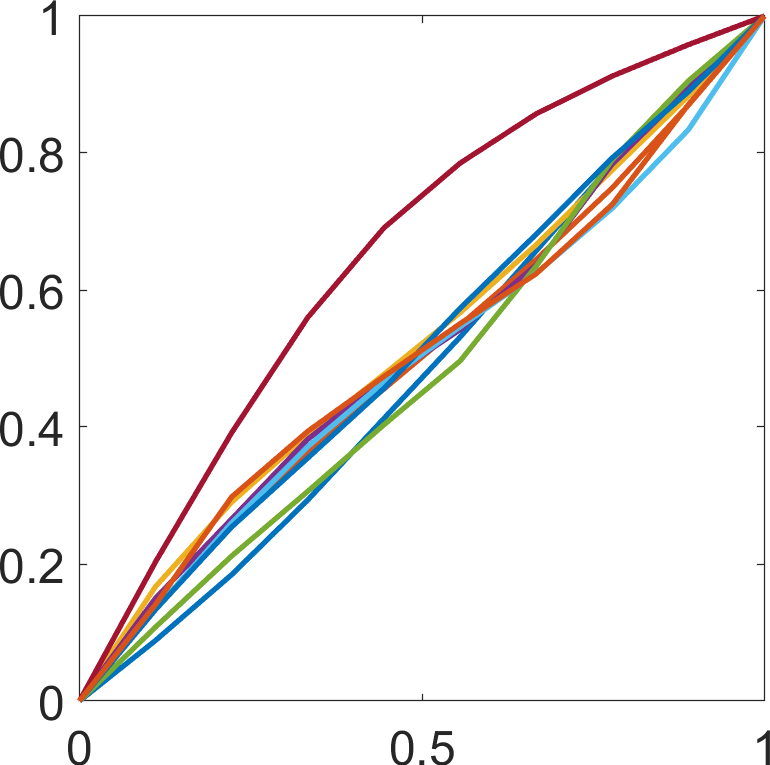}  &  \includegraphics[width = 1.95 in]{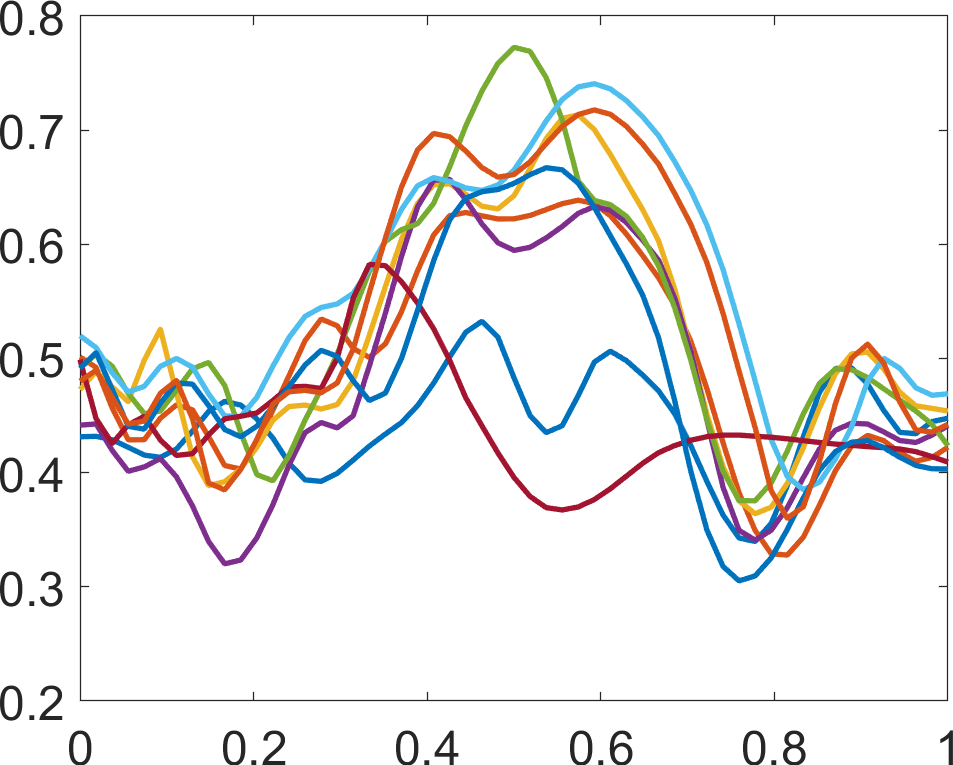} \\
    (d) & (e) & (f)\\
    \end{tabular}
    \caption{(a) Observed complete FA functions. (b) Observed fragmented FA functions. (c) Zoom-in on the fragmented region. Posterior mean estaimtes of amplitude (d) and phase (e) for the nine fragmented FA functions shown in (a). (f) Composition of the amplitude and phase posterior means.}
    \label{FA_data}
    \end{center}
\end{figure}

Diffusion Tensor-Magnetic Resonance Imaging (DT-MRI) is a neuroimaging modality that traces the diffusion of water molecules in the brain. A scan of a subject's brain provides a $3\times3$ matrix at each voxel in the image that describes the constraints of local motion of water molecules. This information is essential to understanding white matter in the brain, which constitutes areas made up of axons or tracts. Tracts connect neurons and allow for the transmittance of electric signals from one area of the brain to another, affecting overall brain function. Due to anisotropic diffusion of water in tracts, they can be extracted from the information contained in a DT-MRI, along with other quantities of interest that describe the quality of a tract connection by summarizing its degree of anisotropy.

Functional anisotropy (FA) measurements along tracts provide a voxelwise summary of the eigenvalues, denoted by $\nu_1,\ \nu_2,\ \nu_3$, of the diffusion matrices. At each voxel in the image, FA is given by the scalar quantity $FA = \sqrt{\frac{3}{2}}\sqrt{\frac{(\nu_1-\bar\nu)^2 + (\nu_2- \bar \nu)^2 + (\nu_3 - \bar \nu)^2}{\nu_1^2 + \nu^2_2 + \nu^2_3}}$, where $\bar \nu = \frac{\nu_1 + \nu_2 + \nu_3}{3}$. A large FA value indicates a large degree of anisotropy. For practitioners, this summary of a DT-MRI provides a measurement of the quality of neuronal connections between particular regions of interest, and has been found to be a useful quantity to study subjects with various diseases including multiple sclerosis (MS) \cite{goldsmith_fa}. In the MS setting, the autoimmune disease causes lesions and damage to tracts that results in a decrease in FA. Thus, FA can be used as a diagnostic measurement to distinguish between healthy controls and subjects with MS, and to predict cognitive and motor disease outcomes. The data of interest in this case takes a functional form, with the domain of the functions representing locations along tracts.

\begin{figure}[!t]
\begin{center}
\begin{tabular}{cccc}
 \includegraphics[width = 1.95 in]{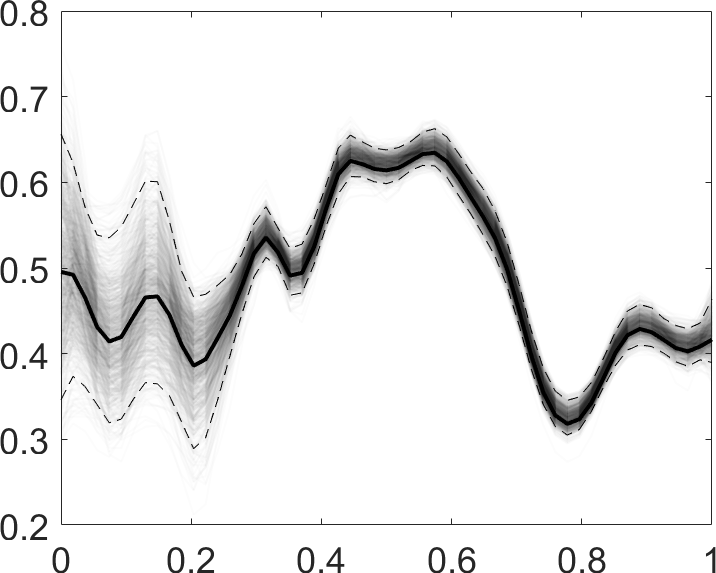} & \includegraphics[width = 1.58 in]{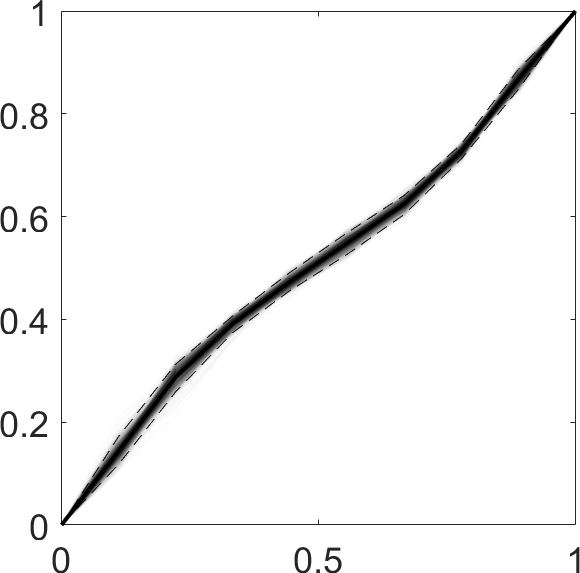}  & \includegraphics[width = 1.95 in]{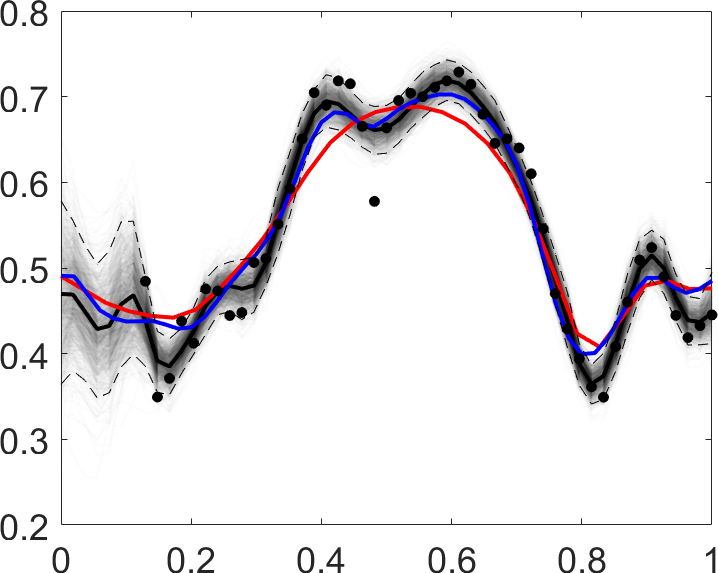} \\
    \includegraphics[width = 1.95 in]{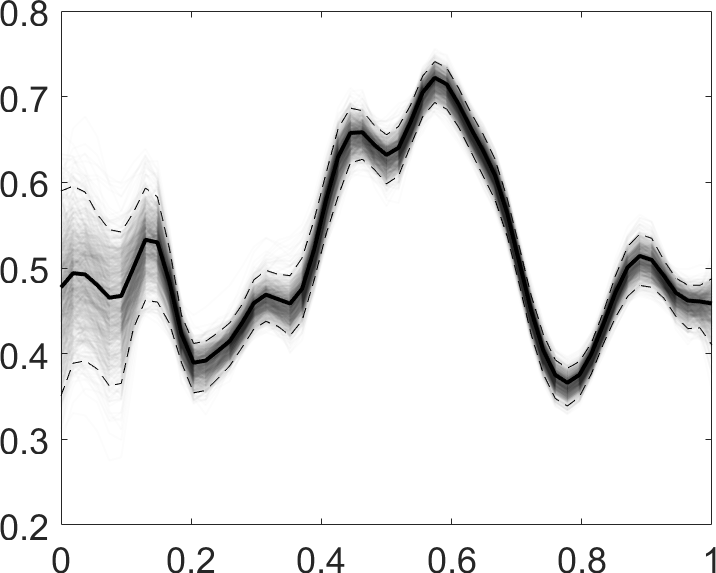} & \includegraphics[width = 1.58 in]{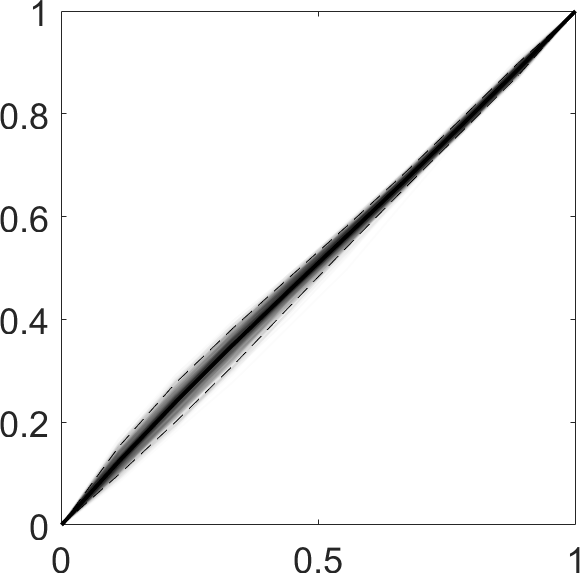} &  \includegraphics[width = 1.95 in]{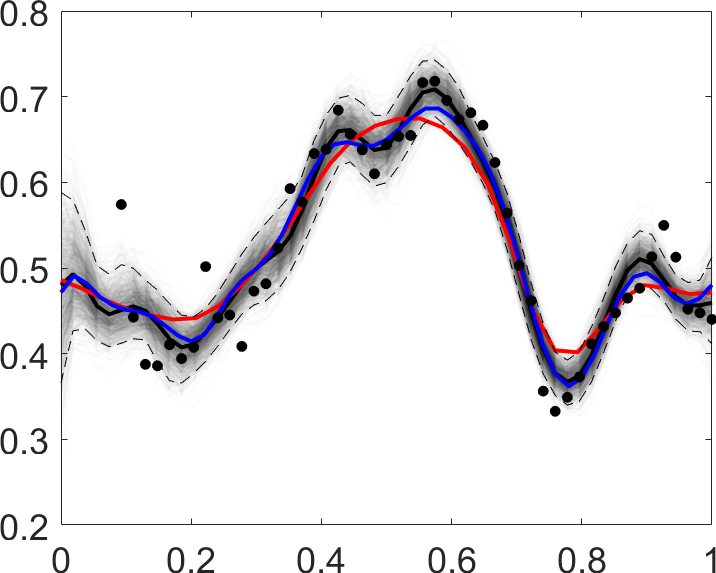}\\
    (a) & (b) & (c) \\
    \end{tabular}
    \caption{Posterior draws (transparent), pointwise mean (solid black) and $95\%$ credible interval (dashed) for the (c) amplitude and (d) phase components. (e) Composition of amplitude and phase, and PACE (red) and WPACE (blue) estimates.}
    \label{FA_examples}
    \end{center}
\end{figure}

A major advantage of DT-MRI as a diagnostic tool is that it is non-invasive. However, FA measurements based on DT-MRI are subject to spurious or missing values due to technical issues with the measuring device. The data that we consider here are 75 FA functions along the right corticospinal tract for subjects diagnosed with MS; the full data is available as part of the 'refund' package in R \cite{goldsmith_refund}. For nine subjects, the recorded FA functions have missing values within the first five tract locations. For the remaining 66 subjects, the FA values were recorded along all 55 tract locations. Figure \ref{FA_data}(a)-(c) shows the 66 complete FA functions, the nine incomplete FA functions, and a zoomed-in view on part of the domain where the fragmentation occurs for improved display, respectively. An empirical amplitude prior is defined based on the 66 complete FA functions. For the remaining settings, we use $\theta_\gamma = .1,\ m_\gamma = 8$ and $B = 16$. Figure \ref{FA_data}(d)-(f) shows marginal posterior mean estimates for amplitude, phase and their composition, respectively, for all nine subjects with fragmented FA functions. All of the estimates generally have similar structure as the complete FA functions. Furthermore, the amplitude functions in (d) provide suitable registration of the estimated FA functions where peaks and valleys, corresponding to different levels of anisotropic diffusion, are well-aligned. Figure \ref{FA_examples} shows two detailed examples of posterior inference for the functional parameters of interest. It also provides posterior summaries, including the posterior mean and $95\%$ credible interval, alongside corresponding PACE and WPACE estimates (an additional example is provided in Figure \ref{motivation_data}). Regardless of the unique features of the different subjects, the proposed Bayesian model provides a reasonable posterior mean function estimate, with much larger uncertainty where the functions are unobserved. There appears to be very little phase uncertainty in these examples. In contrast, the PACE and WPACE estimates appear to severely oversmooth the data.

\subsection{Bayesian Model with Shape-restricted Amplitude Prior}

Next, we focus on examples where the proposed shape-restricted amplitude prior is most appropriate to estimate functions under considerable noise and sparsity.

\subsubsection{Simulated Example 1: Functions with Low Signal-to-Noise Ratio}

\begin{figure}[!t]
\begin{center}
\begin{tabular}{ccc}
 \includegraphics[width = 1.95 in]{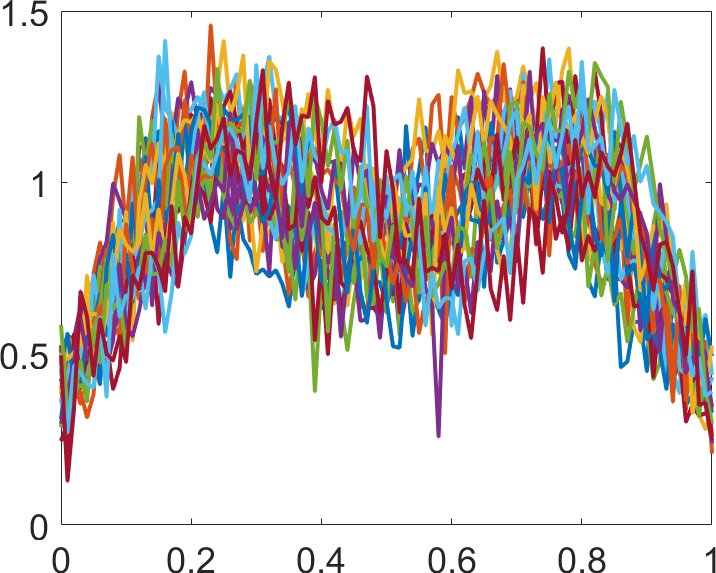} &\includegraphics[width = 1.95 in]{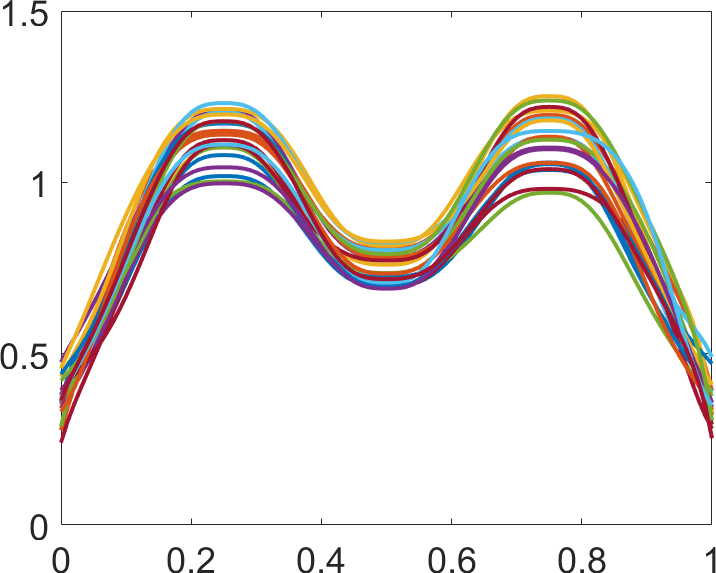} & \includegraphics[width = 1.95 in]{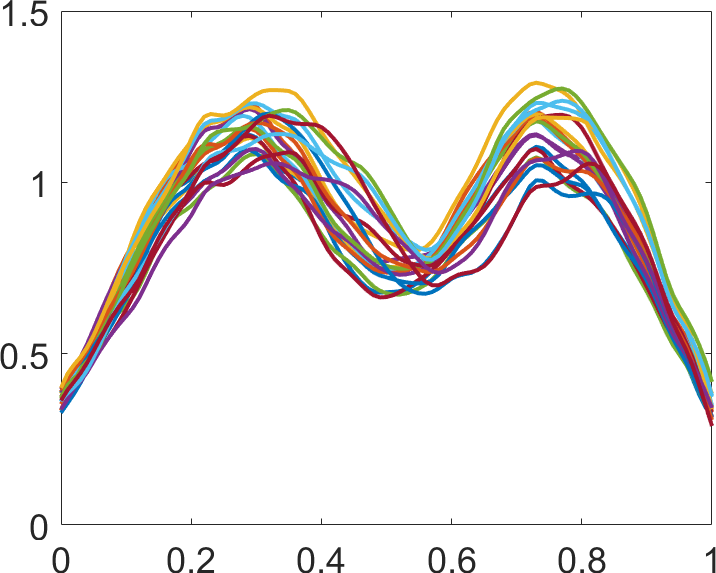}\\
(a)&&\\
  \includegraphics[width = 1.95 in]{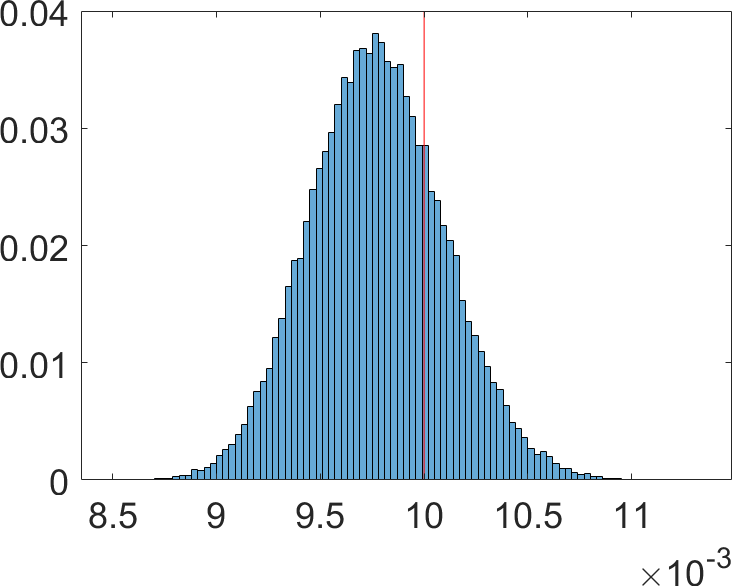}&\includegraphics[width =1.58 in]{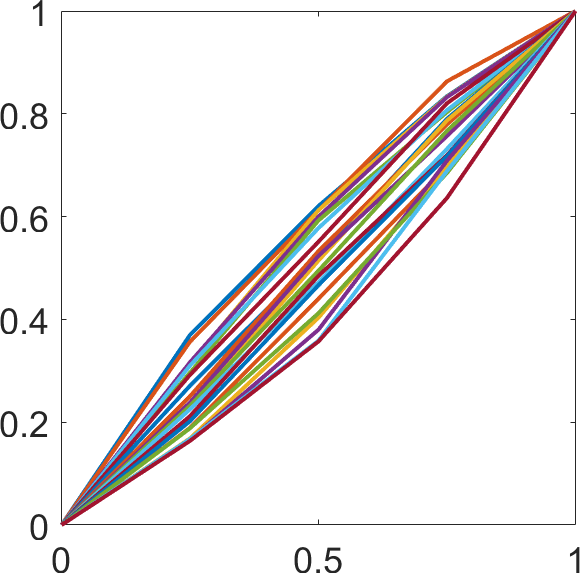} &
 \includegraphics[width = 1.58 in]{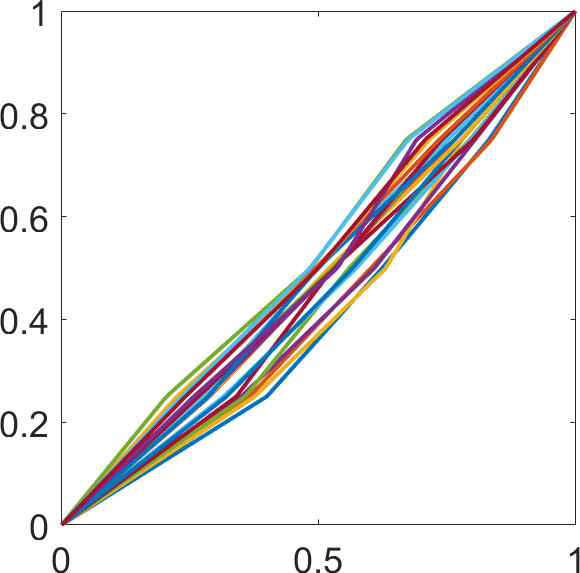} \\
(b) &  (c) & (d)\\
    \end{tabular}
    \caption{(a) Simulated observations with high noise level. (b) Normalized histogram of posterior draws of the error variance $\sigma^2$, with the value used to generate the data in red. (c) Posterior means of the amplitude (top) and phase (bottom) components estimated using the proposed Bayesian model with the shape restricted amplitude prior. (d) Estimated amplitude (top) and phase (bottom) using WPACE.}
    \label{simu_registration}
    \end{center}
\end{figure}

We first consider simulated functional data that not only contains phase and amplitude variability, but also low signal-to-noise ratio. The data is shown in Figure \ref{simu_registration}(a). A shape-restricted amplitude prior is appropriate in this setting since we know that the underlying functions should have two peaks and a single valley. We use the settings $\theta_\gamma = 10,\ m_\gamma = 3,\ B = 10,\ H = 3,\ \alpha = (.25,.5,.75)$ and $M = -1$ in the model. Marginal posterior means of the amplitude and phase components of the unknown functions are shown in the first and second rows of panel (c), respectively. In comparison, the amplitude and phase components estimated by WPACE are given in (d). The WPACE result appears unsatisfactory in two ways: (1) there are significant artifacts in the estimated amplitude functions, and (2) the phase component is underestimated resulting in a fair amount of phase variability that remains in the amplitude estimates. On the other hand, the proposed model is able to appropriately account for the pointwise noise. It results in estimated amplitude functions that have properly aligned peaks and valleys. Furthermore, there is a common degree of smoothness provided by the shape-restricted amplitude prior. In fact, the proposed model is able to estimate the true error variance (red line) in the likelihood fairly well (panel (b)).

\subsubsection{Real Data Example 1: Berkeley Growth Velocity Functions}

\begin{figure}[!t]
\begin{center}
\begin{tabular}{ccc}
 \includegraphics[width = 1.95 in]{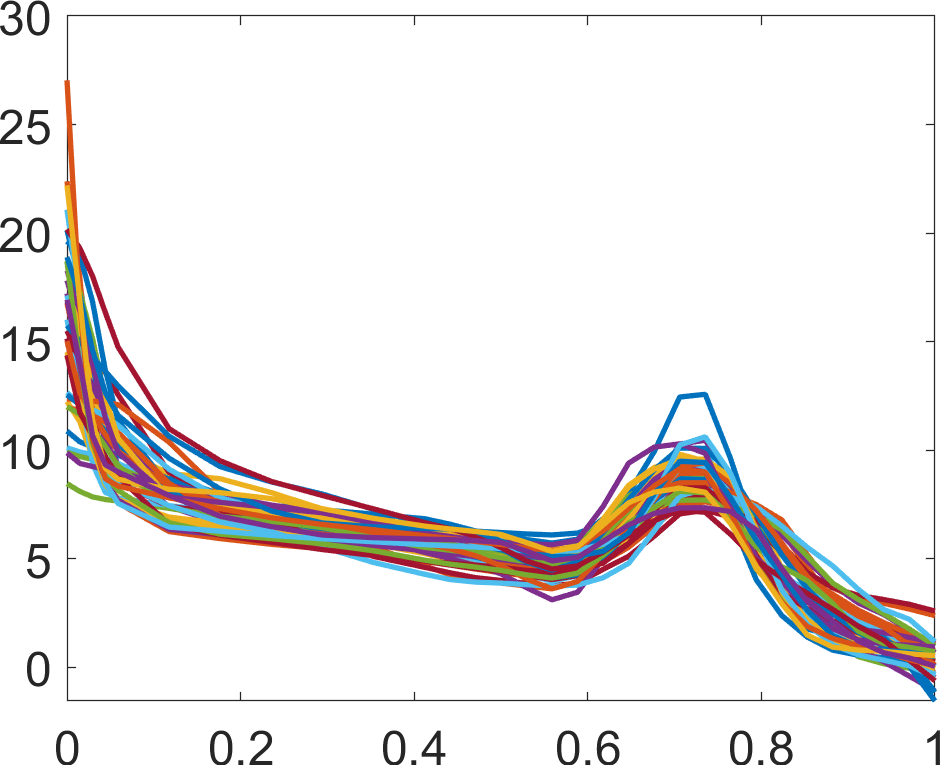} & \includegraphics[width = 1.95 in]{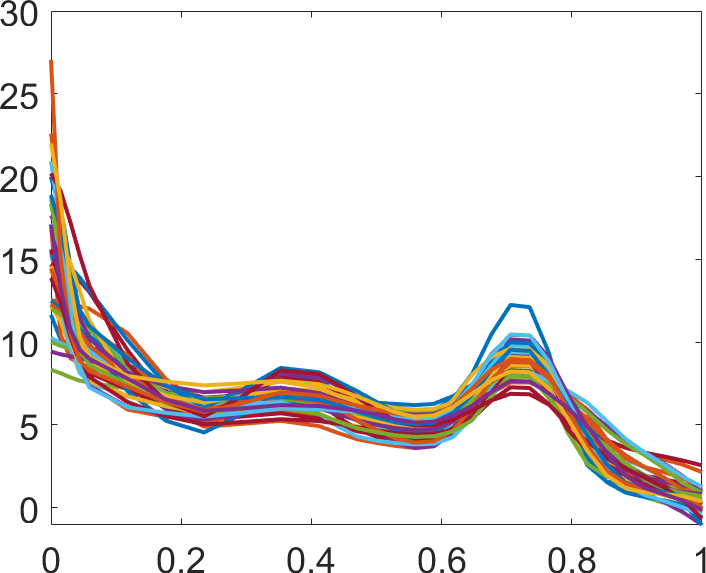}  & \includegraphics[width = 1.95 in]{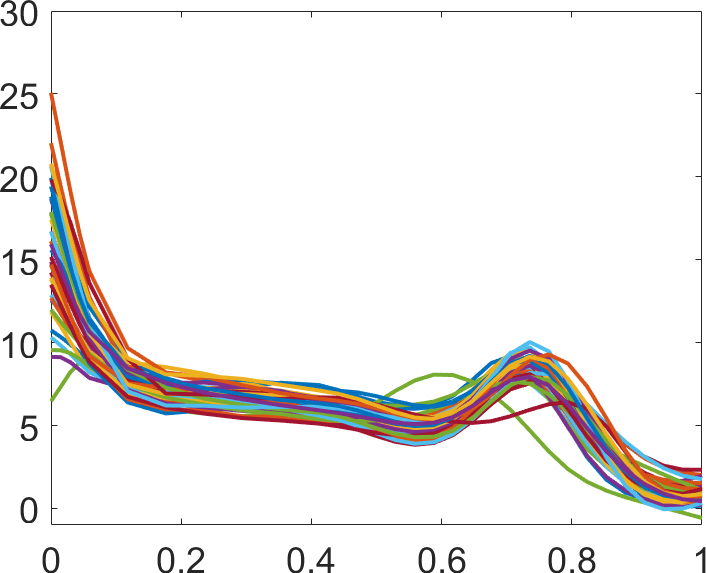} \\
 \includegraphics[width = 1.74 in]{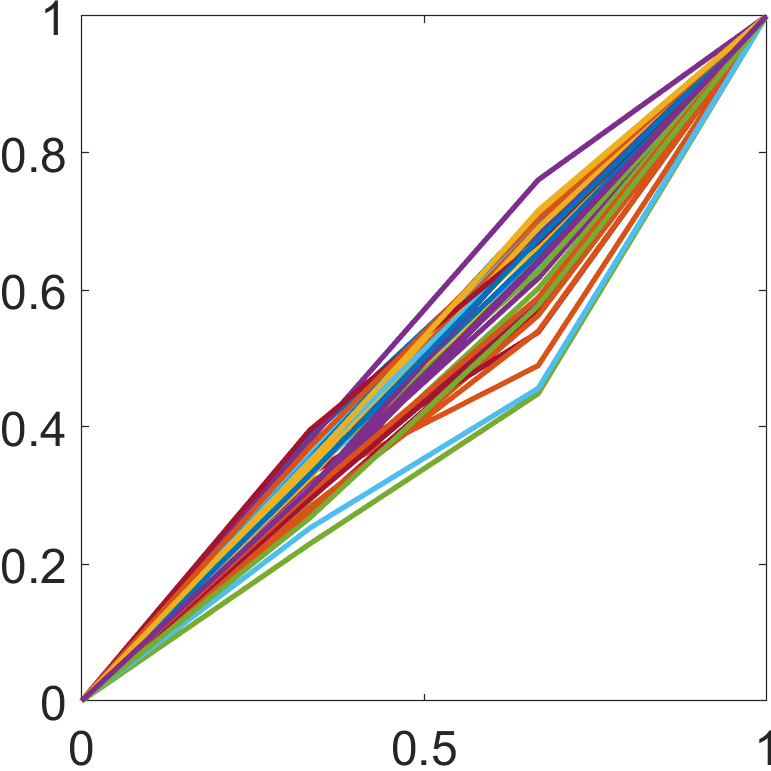}  & \includegraphics[width = 1.75 in]{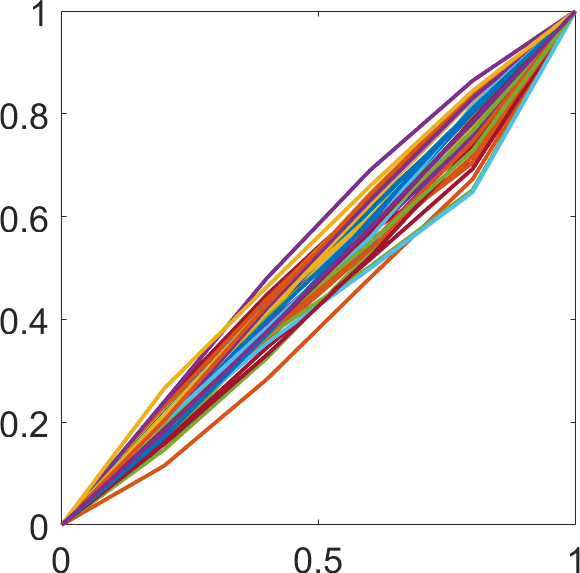}  & \includegraphics[width = 1.75 in]{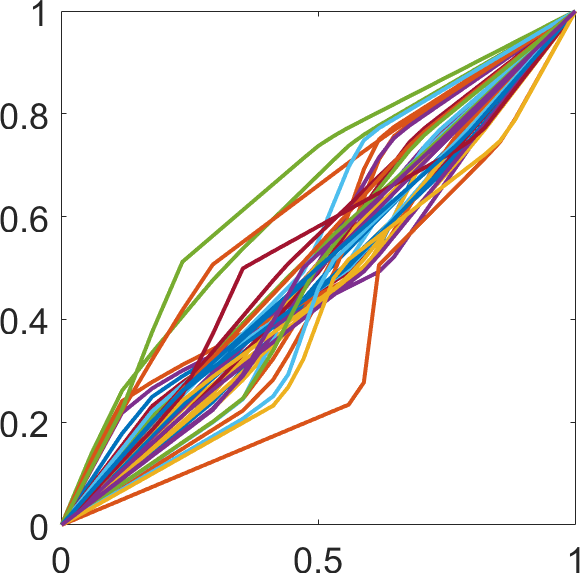} \\
    (b) & (c) & (d) \\
    \end{tabular}
    \caption{Posterior means of the amplitude (top) and phase (bottom) components based on the proposed model with the shape restricted amplitude prior with (a) $H = 2$ and (b) $H = 4$. (c) Estimated amplitude (top) and phase (bottom) functions using WPACE.}
    \label{growth_registration}
    \end{center}
\end{figure}

\begin{figure}[!t]
\begin{center}
\begin{tabular}{ccc}
 \includegraphics[width = 1.95 in]{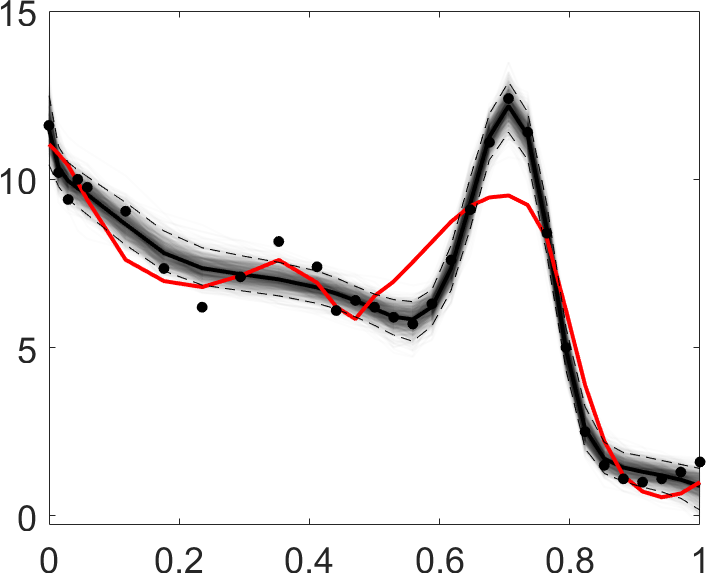} &  \includegraphics[width = 1.95 in]{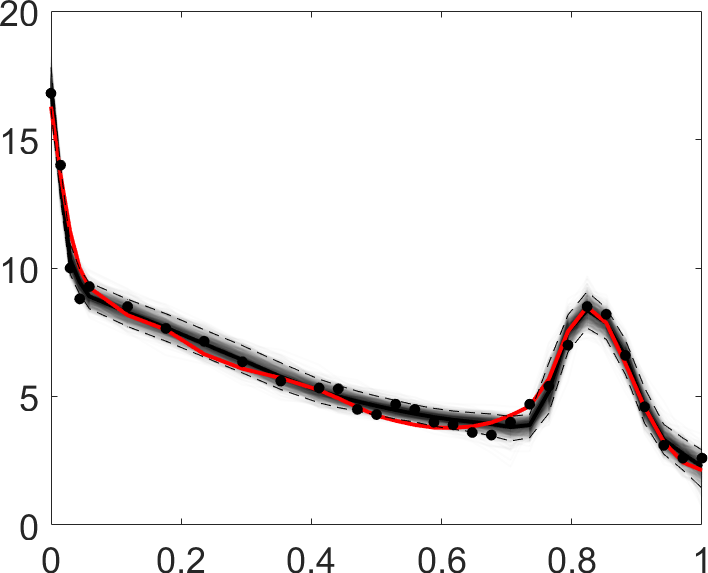}  &  \includegraphics[width = 1.95 in]{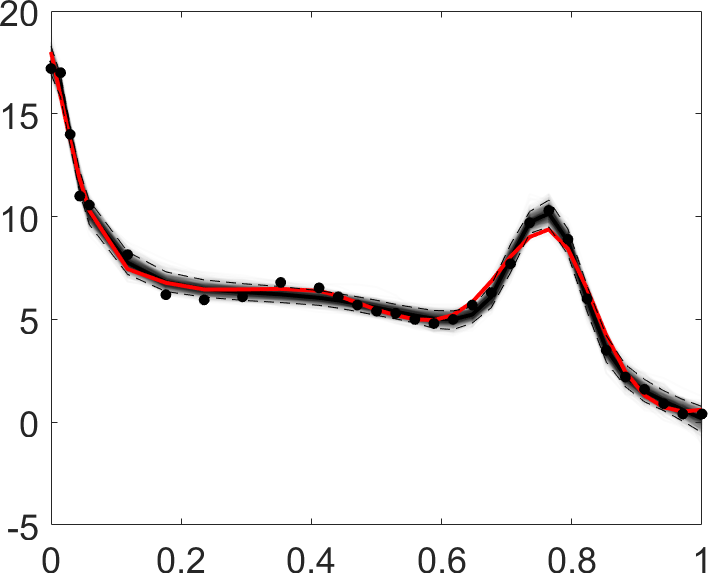} \\
 \includegraphics[width = 1.95 in]{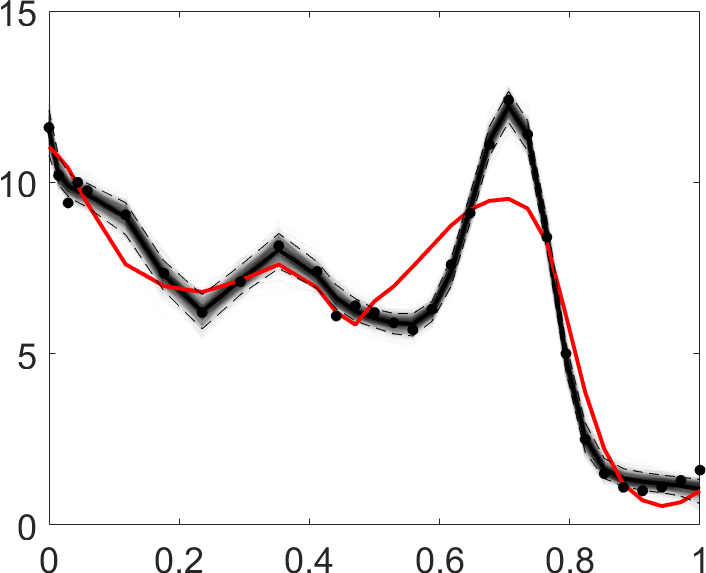} &  \includegraphics[width = 1.95 in]{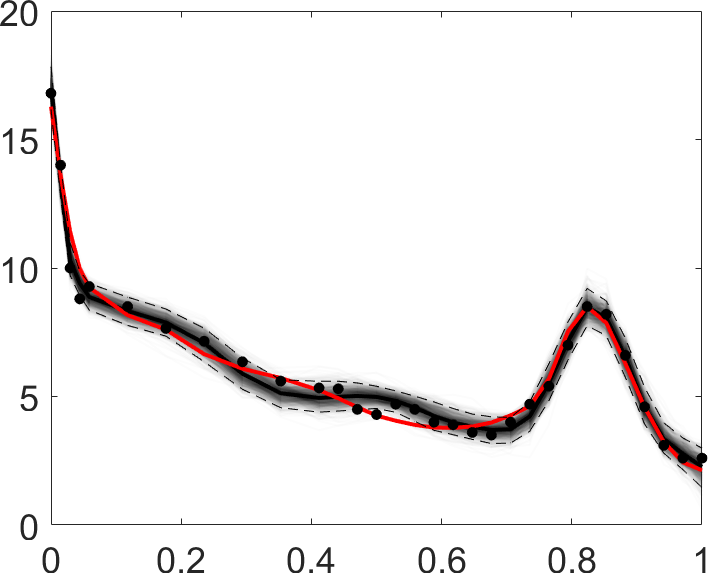}  &  \includegraphics[width = 1.95 in]{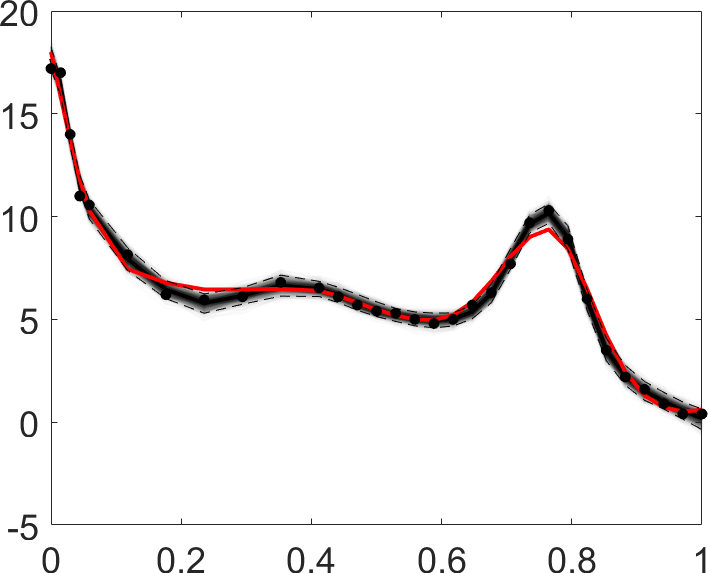} \\
    (a) & (b) & (c) \\
    \end{tabular}
    \caption{Posterior draws (transparent), mean (solid black), and credible interval (dashed) when $H = 2$ and (top) $H = 4$ (bottom), with a comparison to the WPACE fitted function (red) for three different subjects.}
    \label{growth_egs}
    \end{center}
\end{figure}

To further illustrate the structure imposed by modeling amplitude with the shape-restricted prior, we applied our methods to the well-known Berkeley growth dataset \cite{ramsay_FDA}, in which the heights of children were tracked over the course of their lives from one to 18 years of age. In this study, we only use a subset of the data corresponding to 39 boys. In many cases, it is more natural to study the rate or velocity of growth, i.e., the derivative of height with respect to time, than growth itself. This allows for better understanding of growth patterns of the subjects wherein peaks in the velocity functions correspond to growth spurts, with the last, largest peak being the pubertal growth spurt. Consequently, we consider two different settings for our model; the first one allows for a single pubertal growth spurt in the amplitude functions, $H = 2,\ \alpha = (.57,.72)$, while the second one allows for an additional earlier growth spurt, $H = 4,\ \alpha = (.23,.57,.57,.72)$. We also set $\theta_\gamma = 10$ and $B = 20$ in both models.

Figure \ref{motivation_data}(c) shows the observed growth velocity data. The estimated amplitude and phase functions under the two amplitude prior settings are visualized in Figure \ref{growth_registration}(a)-(b). A comparison to the result generated by WPACE is given in panel (c). As expected, the registration and level of smoothness for the pubertal growth spurt is similar across panels (a) and (b). The main difference between the two sets of amplitude estimates is in the additional growth spurt estimated for some of the subjects in (b). Contrasting these results to the WPACE registration and smoothing results shown in (c), the pre-pubertal growth spurt is often smoothed-out, with the model failing to properly align many of the subjects' pubertal growth spurts. This is especially surprising since the phase functions estimated by WPACE are much less regular than those estimated using the proposed model. Figure \ref{growth_egs} shows detailed inferential results for three individuals with and without the pre-pubertal growth spurt. In panel (a), it appears that the subject has a fairly significant pre-pubertal growth spurt. When the amplitude prior in our model is restricted to allow for a single peak, the pattern of observations around this growth spurt is treated as noise and consequently it is missed in the resulting estimated function. Additionally, posterior uncertainty in this region is relatively large. In contrast, when the prior is relaxed to allow for an additional growth spurt, we are able to nicely estimate both the pre-pubertal and the pubertal growth spurts. In panel (b), there appears to be a single large growth spurt. Both estimates provided by our model appear to fit the data well. Finally, in (c), it is unclear whether there is a small pre-pubertal growth spurt. Again, both estimates provided by our model are reasonable. This example suggests that fixing the amplitude hyperparameter $H$ a-priori can be limiting in practice and motivates future work to jointly estimate $H$ for different subjects. The WPACE estimates are shown in each panel in red; the WPACE estimate in (a) appears to severely oversmooth and underestimate the pubertal growth spurt.

\subsubsection{Real Data Example 2: Bone Mineral Density}

\begin{figure}[!t]
\begin{center}
\begin{tabular}{cc}
\includegraphics[width = 2 in]{figures_whitespace/BMD_temp.png}&\includegraphics[width = 2 in]{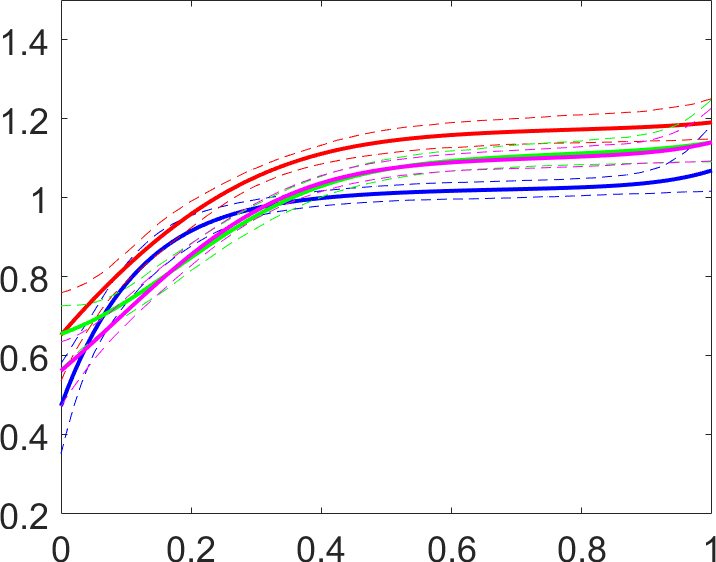}\\
(a)&(b)\\
    \end{tabular}
\caption{(a) BMD measurements colored by ethnicity: Asian (blue), Black (red), Hispanic (green) and Caucasian (magenta). (b) Posterior estimates of mean BMD functions, with $95\%$ credible intervals, for the four ethnic groups.}
    \label{BMD_fig}
    \end{center}
\end{figure}

As a last example, we consider indirect x-ray measurements of bone mineral density (BMD), associated with skeletal health and diseases such as osteoporosis \cite{bachrach_BMD}. While osteoporosis affects individuals later in life, bone development during adolescence through early adulthood can be used to assess an individual's risk for the disease. We focus on a subset of the entire dataset corresponding to females aged nine to 25 years that had their BMD measured during two, three or four doctor appointments over the course of several years. Of primary interest in the study was the mean difference in BMD for four different ethnicities. Figure \ref{BMD_fig}(a) shows the data where the different ethnic groups are highlighted using different colors. While this subset of the data has been used to classify individuals in previous work \cite{james_BMD,delaigle_BMD}, our focus is on estimation of a mean BMD trajectory for each group that accommodates phase variability. We restrict our model by forcing all individuals within the same ethnic group to have a common translation and a common strictly increasing amplitude function (prior hyperparameters are set to $B = 5,\ H = 0$ and $M=1$). We do allow for individual phase variability with $\theta_\gamma = 100$ and $m_\gamma = 1$, resulting in very simple and regular phase function estimates. The composition of the amplitude and phase samples then corresponds to individual BMD trajectories, and our interest is in their mean estimate. Our assumption that the mean BMD function is strictly increasing stems from the age range of the individuals in the study, and enforces the principle that, on average, BMD increases during this period.

The resulting groupwise mean BMD trajectory estimates, with $95\%$ credible intervals, are shown in Figure \ref{BMD_fig}(b). The results obtained via the proposed model visually corroborate the major finding of the original study that the Black ethnic group (red) has a significantly higher BMD than the other groups. Structurally, the estimated mean function for the Asian group, shown in blue, is also quite different from the others: there is a lot of BMD growth early followed by minimal growth later on in life. The estimated mean BMD functions for the Caucasian and Hispanic groups appear extremely similar. The original study concluded that the Asian, Caucasian and Hispanic groups have BMD patterns that are difficult to distinguish from one another.




\section{Discussion and Future Work}
\label{conclusion}

We have presented a Bayesian framework for simultaneous registration and inference for functional observations that can handle challenging observational regimes such as sparsity, fragmentation and low signal-to-noise ratio. The framework explicitly accounts for phase and amplitude variability, and imposes amplitude restrictions in situations where different types of prior information are available. Indeed, we show that the key to inference under general observation regimes is the ability to inform the prior on the underlying structure of amplitude. We demonstrate the performance of the proposed model on several different simulated and real data examples to show the diverse range of scenarios that can be analyzed. We also compare the proposed method to a state-of-the-art competitor.

We have identified two main directions for future work. First, in both the empirical and shape-restricted amplitude prior scenarios, we suggest additional modeling strategies that will further improve the proposed framework. In the first case, we will incorporate an additional hierarchical layer corresponding to subspace estimation (via fPCA) for the amplitude subspace. This will allow for direct propagation of uncertainty from the training stage to the estimation stage. In the second case, we will treat the number of local extrema allowed in the amplitude estimates $H$ as well as their ordering pattern $M$, as unknown quantities to be estimated. This will result in a shape-restricted amplitude mixture prior wherein each mixture component corresponds to a different combination of $H$ and $M$. Finally, we will extend the proposed model to other functional data scenarios including shapes of higher-dimensional curves, images and shapes of surfaces.

\noindent\textbf{Acknowledgments:} We acknowledge the Isaac Newton Institute for Mathematical Sciences for hosting the Variational Methods and Effective Algorithms for Imaging and Vision in Fall 2017 where many of the research ideas presented in this article originated. This research was partially funded by NSF DMS-1613054, NIH R37-CA214955 (to SK and KB), and NSF CCF-1740761 and NSF CCF-1839252 (to SK).

\bibliographystyle{siam}
\bibliography{bibliography}
\end{document}